\documentclass{aa} 
\usepackage[breaklinks=true, colorlinks=true, linkcolor=blue, citecolor=blue]{hyperref}
\usepackage{graphicx}
\DeclareGraphicsExtensions{.pdf,.png,.jpg,.ps,.eps}
\usepackage{xcolor}
\usepackage{ltablex}
\usepackage{placeins}
 \usepackage{lscape}
\usepackage{color}

\usepackage[varg]{txfonts}

\begin{document}

   \title{New massive members of Cygnus OB2}

   \author{S. R. Berlanas
          \inst{1,2}
         \and
          A. Herrero\inst{1,2}
          \and
          F. Comerón\inst{3}
          \and
          A. Pasquali\inst{4}
          \and
          C. Bertelli Motta \inst{4}
          \and
          A. Sota\inst{5}   
         }

   \institute{Instituto de Astrofísica de Canarias, 38200 La Laguna, Tenerife, Spain\   			              
         \and Departamento de Astrofísica, Universidad de La Laguna, 38205 La Laguna, Tenerife, Spain\	            
         \and ESO, Karl-Schwarzschild-Strasse 2, 85748 Garching bei München, Germany\ 
         \and Astronomisches Rechen-Institut, Zentrum für Astronomie der Universität Heidelberg, Mönchhofstr 12–14, 69120 Heidelberg, Germany\   
          \and Instituto de Astrofísica de Andalucía-CSIC, 18008 Granada, Spain\      
             }

   \date{Received month day, year; accepted month day, year}

  \abstract
   { The Cygnus complex is one of the most powerful star forming regions at a close distance from the Sun ($\sim$1.4 kpc). Its richest OB association Cygnus OB2 is known to harbor many tens of O-type stars and hundreds of B-type stars, providing a large homogeneous population of OB stars that can be analyzed. Many studies of its massive population have been developed in the last decades, although the total number of OB stars is still incomplete.}
   {Our aim is to increase the sample of O and B members of Cygnus OB2 and its surroundings by spectroscopically classifying 61 candidates as possible OB-type members of Cygnus OB2, using new intermediate resolution spectroscopy.}
   {We have obtained intermediate resolution (R$\sim$5000) spectra for all of the OB-type candidates between 2013 and 2017. We thus performed a spectral classification of the sample using  HeI-II and metal lines rates, as well as the Marxist Ghost Buster (MGB) software for O-type stars and the IACOB standards catalog for B-type stars.}
   {From the whole sample of 61 candidates, we have classified 42 stars as new massive OB-type stars, earlier than B3, in Cygnus OB2 and surroundings, including 11 O-type stars. The other candidates are discarded as they display later spectral types inconsistent with membership in the association. We have also obtained visual extinctions for all the new confirmed massive OB members, placing them in a Hertzsprung-Russell Diagram using calibrations for $T_{\rm eff}$ and luminosity. Finally, we have studied the age and extinction distribution of our sample within the region. }
   { We have obtained new blue intermediate-resolution spectra suitable for spectral classification of 61 OB candidates in Cygnus OB2 and surroundings. The confirmation of 42 new OB massive stars (earlier than B3) in the region allows us to increase the young massive population known in the field. We have also confirmed the correlation between age and Galactic longitude previously found in the region. We conclude that many O and early B stars at $B>16$ mag are still undiscovered in Cygnus.}


   \maketitle
%

\section{Introduction}

The Cygnus region is the most powerful nearby stellar complex, conspicuous at all wavelengths and very young, with several rich OB associations, numerous young open clusters and tens of compact H\,II and star formation regions in the field. Hosting the largest number of nearby massive stars and an intense star forming activity \citep{reipurth08}, it provides an updated view of the high-mass stellar population in one of the largest groups of young stars in our Galaxy. It is an ideal place to study the process of massive star formation and evolution, individually and in stellar groups, and their interaction with the surroundings.

Its association Cygnus OB2 \citep[d$\sim$1.4 kpc,][]{rygl12}  has received a lot of attention and has been studied at all wavelengths with different spatial coverage since it hosts a high number of early spectral type stars \citep{wal02}. First studies were carried out by  \cite{morgan54}, \cite{schulte56,schulte58} and \cite{reddish68}, but it was \cite{mt91} who developed an extensive survey  of the massive population in the association, identifying 120 possible massive star members, 70 of which were classified as OB stars (42 O-type stars). \cite{knodlseder00} proposed that this number should be much larger, around 100 O-type stars. In the last few decades, many other studies were carried out in the region updating continually this number \citep{com02,hanson03,kiminki07,com08,negueruela08, com12}. These surveys have allowed the global study of the massive population in the region, using photometry to place the stars in a Hertzprung-Russell diagram from which the star formation history and mass function of the association were assessed \citep{wright15}.
In spite of the many photometric and spectroscopic surveys carried out in the region, only a small  homogeneous group of early type stars have been spectroscopically analyzed \citep{herrero99,herrero02,negueruela08}, and few stars have been observed in the UV range \citep{herrero01}.
There is still a large number of stars that should be explored.
The optical extinction of the region is high \citep[$A_{\rm V}= 4.0-7.0$ mag,][]{wright15}, but not so much as to prevent obtaining spectra of its most massive stars for a rough spectral classification.
Therefore, new spectroscopy to search for previously undiscovered massive stars is mandatory to complete the last census of massive O and B-type stars in the association. 

One of the most complete spectroscopic surveys in the Cygnus region was developed by \cite{com12}. They performed spectral classification of a magnitude-limited sample ($B\leq$ 16 mag and $Ks<9$ mag) selected with a homogeneous photometric criterion over a large area that includes Cygnus OB2 and its surroundings, providing a large sample of known and new OB stars, as well as a list of 61 OB candidates for which no spectral data is available and that are pending spectroscopic confirmation. 

The main goal of this work is to complete the spectral classification of this latter sample, aiming a later determination of the stellar parameters.
Thus, we have obtained intermediate-resolution spectra of all the list candidates, in order to confirm or reject them as true massive OB-type stars.
In Sect.~\ref{sect2} we present the observations. In Sect.~\ref{sect3} we describe the spectral classification criteria used in this work, and in Sect.~\ref{sect4} visual extinctions and stellar parameters derived for the new OB-type members. The results are discussed in Sect.~\ref{sect5} where we show the Hertzsprung-Russell diagram (HRD) of the region and the age distribution found across Galactic longitude. Finally, we summarize our conclusions in Sect.~\ref{sect6}.

\section{Observations and data reduction}\label{sect2}

The study developed by \cite{com12} produced a sample of O and early B stars (in a 6$^{\circ}$ x 4$^{\circ}$ region centered on Galactic coordinates $l=$ 79.8$^{\circ}$ and $b=$ +0.8$^{\circ}$ of Cygnus OB2) which were identified using two homogeneous reddening-free criteria
\begin{align}
Q_{BJK} = 0.196 (B-J)-0.981 (J-K)-0.098 > 0\\ 
Q_{JHK} = 0.447 (J-H)-0.894 (H-K)-0.089 < 0
\end{align}
\noindent which allowed them to classify 60 new OB stars and produce a list of 61 candidates pending spectroscopic data. 
They used $BJHK$ photometry tabulated in the USNO-B \citep{USNO04,USNO10} and 2MASS all-sky \citep{2MASS} catalogs setting limiting magnitudes of $B\leq16$ mag and $Ks<9.0$ mag. By combining these two magnitude cuts with the $(B-K)$ colors of OB stars, \cite{com12} set up a selection method sensitive to main-sequence stars earlier than B1 and obscured by $A_{\rm V}<6.7$ mag.

\begin{figure}[ht!]
\centering
\includegraphics[width=8.8cm]{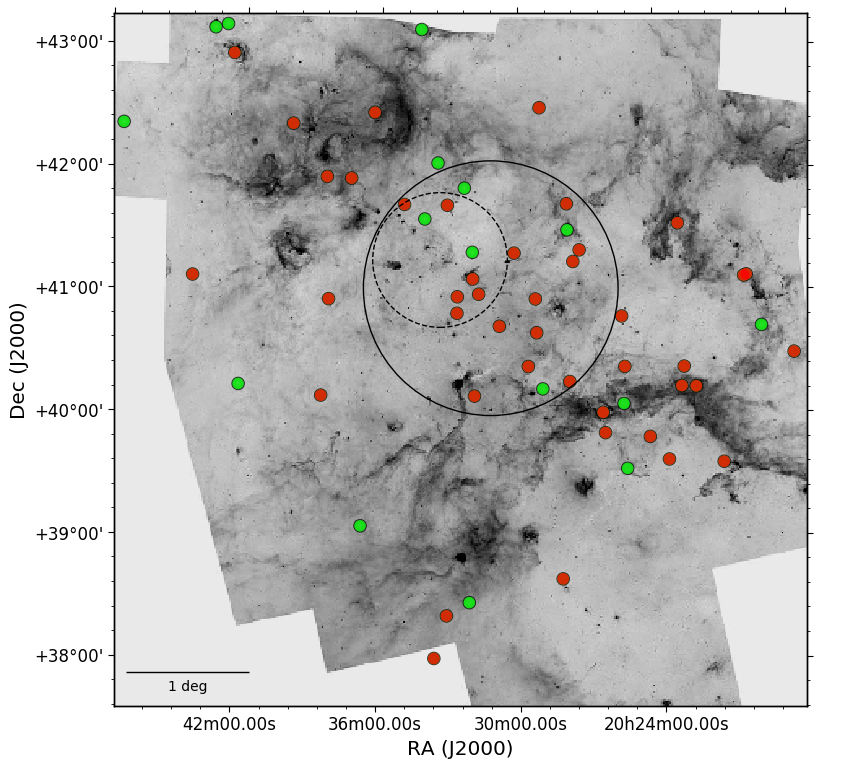}
\caption{Inverse Spitzer 8 $\mu$m image of the Cygnus region showing the location of the 61 OB candidates. The 42 confirmed massive OB-type stars earlier than B3 are indicated with red dots. The remaining stars are late B and foreground A-F-G stars which are indicated with green dots. The solid line circle delimits the 1 degree radius area of Cygnus OB2 adopted by \cite{com12}. For reference, the dash-dotted line circle shows the area considered by \cite{wright15}.}
\label{fig1}
\end{figure}

We have obtained new spectra for all the proposed OB candidates, whose location is shown in Fig.~\ref{fig1}. The sample has been observed in five different runs between 2013 and 2017. For an accurate spectral classification of OB-type stars we need blue spectra (4000-5000 $\AA$) where the diagnostic He\,I-II and metal lines are located. 

\begin{table}[b!]
\centering
\caption{Telescopes, instruments and settings used in this work.}	
		\label{table1}
		\begin{tabular}{lcccc}
		\hline 
		\hline \\[-1.5ex]  
   		 \small{Instrument}& \small{Telescope}& \small{Resol.} & \small{Date}& \small{Stars} \\
   		\small{and grating}& \small{}& \small{} & \small{}& \small{} \\ 
   		 
   		\hline\\[-1.5ex]  
   		\small{ISIS - 600B} & \small{ORM-WHT} & \small{2500}& \small{Oct 2013}& \small{1} \\
   		\small{WYFFOS - H2400B} & \small{ORM-WHT}  & \small{5000}& \small{Jul 2014}& \small{7} \\
   		\small{ISIS - H2400} & \small{ORM-WHT}  & \small{7500}& \small{Jul 2015}& \small{2} \\  		
   		\small{OSIRIS - R2500U/V} & \small{ORM-GTC}  & \small{2500}& \small{May 2016}& \small{2} \\	
      	\small{IDS - R1200B } & \small{ORM-INT} & \small{5000}& \small{Jul 2016}& \small{47}\\
 		\small{ISIS - R1200B} & \small{ORM-WHT}& \small{5000}& \small{Apr 2017}& \small{2} \\
      	\hline
		\end{tabular}
				
\end{table}

The bulk of stars were observed in July 2016, obtaining spectra of 47 candidates. We have chosen the R1200B grating of IDS (EEV10) on the Isaac Newton Telescope (INT) in La Palma, which provides a resolution of $\sim$5000 at 4500 $\AA$. The remaining stars were observed in different runs between 2013 and 2017, at the William Herschel Telescope (WHT) using the AF2/WYFFOS and ISIS instruments, and at the Gran Telescopio CANARIAS (GTC) using the OSIRIS instrument. Out of the whole sample, three of the stars (J20301097+4120088, J20315433+4010067, J20345785+4143543) belong to the Galactic O-Star Spectroscopic Survey (GOSSS) catalog \citep{maiz16}.

The information on the different runs carried out in this work is shown in Table~\ref{table1}. The final spectra were reduced using the IRAF procedures, with standard routines for bias and flat-field subtraction and also for the wavelength calibration.

\section{Spectral classification}\label{sect3}

 \begin{table*}[t!]
 \centering
 \caption{Basic data of the confirmed massive new members (earlier than B3).}
 \label{table2}
		\begin{tabular}{lcccccccc}
		\hline 
		\hline \\[-1.5ex]  
    	\small{Object}&\small{RA (hhmmss)}& \small{Dec ($^{\circ}$ $^\prime$ $^{\prime\prime}$)}& \small{Region}& \small{$B$} & \small{$Ks$} & \small{$J$} & \small{SpT}& \small{Binary star}\\

    \hline \\[-1.5ex]
    
\small{J20423509+4256364} & \small{20 42 35.08} & \small{+42 56 36.43}& \small{\textit{c}}& \small{14.480}& \small{8.304}& \small{9.211}& \small{O6IIIz}& \small{}  \\     
\small{J20371773+4156316} & \small{20 37 17.73} & \small{+41 56 31.57}& \small{\textit{c}}& \small{15.760}& \small{8.041}& \small{9.071}& \small{O7V}& \small{}  \\     
\small{J20345785+4143543} & \small{20 34 57.84} & \small{+41 43 54.25}& \small{\textit{a}}& \small{15.430}& \small{7.417}& \small{8.447}& \small{O7:Ib}& \small{}  \\    
\small{J20293563+4024315} & \small{20 29 35.63} & \small{+40 24 31.45}& \small{\textit{b}}& \small{12.468}& \small{8.268}& \small{8.831}& \small{O8IIIz}& \small{} \\    
\small{J20222481+4013426} & \small{20 22 24.81} & \small{+40 13 42.55}& \small{\textit{c}}& \small{13.620}& \small{8.410}& \small{9.034}& \small{O8II}& \small{}  \\ 
\small{J20261976+3951425} & \small{20 26 19.75} & \small{+39 51 42.46}& \small{\textit{c}}& \small{15.600}& \small{8.348}& \small{9.351}& \small{O8.5IV}& \small{} \\ 
\small{J20275292+4144067} & \small{20 27 52.92} & \small{+41 44 06.65}& \small{\textit{b}}& \small{13.330}& \small{7.277}& \small{8.144}&  \small{O9.5II}& \small{} \\ 
\small{J20262484+4001413} & \small{20 26 24.84} & \small{+40 01 41.25}& \small{\textit{c}}& \small{13.290}& \small{8.330}& \small{9.021}& \small{O9.2III}& \small{} \\
\small{J20291617+4057372} & \small{20 29 16.17} & \small{+40 57 37.19}& \small{\textit{b}}& \small{15.030}& \small{7.899}& \small{8.855}& \small{O9.7III} & \small{} \\ 
\small{J20382173+4157069} & \small{20 38 21.72} & \small{+41 57 06.89}& \small{\textit{c}}& \small{15.810}& \small{7.682}& \small{8.760}& \small{O9.7II} & \small{} \\ 
\small{J20181090+4029063} & \small{20 18 10.89} & \small{+40 29 06.29}& \small{\textit{c}}& \small{14.940}& \small{8.399}& \small{9.343} & \small{O9.7Ib}& \small{yes*} \\ 
\small{J20273787+4115468} & \small{20 27 37.87} & \small{+41 15 46.79}& \small{\textit{b}}& \small{14.570}& \small{8.263}& \small{9.146}& \small{B0II} & \small{}\\
\small{J20301097+4120088} & \small{20 30 10.97} & \small{+41 20 08.82}& \small{\textit{b}}& \small{15.690}& \small{8.882}& \small{9.855}& \small{B0:II:}& \small{}  \\ 
\small{J20323968+4050418} & \small{20 32 39.68} & \small{+40 50 41.83}& \small{\textit{a}}& \small{14.410}& \small{8.913}& \small{9.631}& \small{B0II}& \small{}   \\ 
\small{J20395358+4222506} & \small{20 39 53.58} & \small{+42 22 50.62}& \small{\textit{c}}& \small{15.890}& \small{5.822}& \small{7.345}& \small{B0I} & \small{yes*} \\
\small{J20281176+3840227} & \small{20 28 11.75} & \small{+38 40 22.73}& \small{\textit{c}}& \small{11.805}& \small{7.944}& \small{8.349}& \small{B0Ib}& \small{}  \\
\small{J20323882+4058469} & \small{20 32 38.82} & \small{+40 58 46.85}& \small{\textit{a}}& \small{15.430}& \small{8.821}& \small{9.701}& \small{B0Ib} & \small{} \\ 
\small{J20225451+4023314} & \small{20 22 54.50} & \small{+40 23 31.39}& \small{\textit{c}}& \small{13.107}& \small{8.601}& \small{9.175}& \small{B0Iab}& \small{} \\
\small{J20253320+4048444} & \small{20 25 33.19} & \small{+40 48 44.38}& \small{\textit{c}}& \small{13.112}& \small{7.648}& \small{8.340}& \small{B0Iab}& \small{} \\
\small{J20272099+4121262} & \small{20 27 20.99} & \small{+41 21 26.15}& \small{\textit{b}} & \small{13.830}& \small{8.730}& \small{9.448} & \small{B0.5V}& \small{yes}\\
\small{HDE229258} & \small{20 24 25.51} & \small{+39 49 28.30}& \small{\textit{c}}& \small{10.235}& \small{8.689}& \small{8.833} & \small{B0.7V}& \small{}\\ 
\small{J20330526+4143367} & \small{20 33 05.26} & \small{+41 43 36.74}& \small{\textit{a}}& \small{13.940}& \small{8.634}& \small{9.286} & \small{B0.5III}& \small{}\\ 
\small{J20361806+4228483} & \small{20 36 18.06} & \small{+42 28 48.30}& \small{\textit{c}}& \small{15.650}& \small{8.855}& \small{9.814} & \small{B0.7III}& \small{}\\ 
\small{J20233816+3938118} & \small{20 23 38.16} & \small{+39 38 11.84}& \small{\textit{c}}& \small{11.293}& \small{8.731}& \small{8.996} & \small{B0.7Ib}& \small{}\\  
\small{HD228973} & \small{20 20 07.35} & \small{+41 07 46.72}& \small{\textit{c}}& \small{10.34}& \small{7.684}& \small{7.922} & \small{B1V}& \small{yes}\\ 
\small{J20201435+4107155} & \small{20 20 14.34} & \small{+41 07 15.45}& \small{\textit{c}}& \small{12.412}& \small{8.240}& \small{8.627} & \small{B1V}& \small{}\\    
\small{J20230290+4133466} & \small{20 23 02.90} & \small{+41 33 46.59}& \small{\textit{c}}& \small{14.780}& \small{7.809}& \small{8.762} & \small{B1V}& \small{}\\ 
\small{BD+404193} & \small{20 29 13.55} & \small{+40 41 03.38}& \small{\textit{b}}& \small{10.412}& \small{8.812}& \small{8.945} & \small{B1V}& \small{}\\ 
\small{BD+404208} & \small{20 30 49.97} & \small{+40 44 18.53}& \small{\textit{b}}& \small{10.654}& \small{8.664}& \small{8.869} & \small{B1V}& \small{}\\
\small{J20314341+4100021} & \small{20 31 43.40} & \small{+41 00 02.07}& \small{\textit{a}}& \small{15.940}& \small{8.957}& \small{9.885} & \small{B1V}& \small{yes*}\\
\small{J20315898+4107314} & \small{20 31 58.98} & \small{+41 07 31.41}& \small{\textit{a}}& \small{15.490}& \small{8.832}& \small{9.773} & \small{B1V}& \small{yes*} \\
\small{J20330453+3822269} & \small{20 33 04.53} & \small{+38 22 26.91}& \small{\textit{c}}& \small{11.383}& \small{8.790}& \small{9.021} & \small{B1V}& \small{}\\
\small{J20230183+4014029} & \small{20 23 01.83} & \small{+40 14 02.90}& \small{\textit{c}}& \small{13.465}& \small{8.060}& \small{8.579} & \small{B1III}& \small{}\\
\small{J20274925+4017004} & \small{20 27 49.25} & \small{+40 17 00.42}& \small{\textit{b}}& \small{13.460}& \small{8.104}& \small{8.713} & \small{B1III}& \small{}\\
\small{J20315433+4010067} & \small{20 31 54.33} & \small{+40 10 06.71}& \small{\textit{b}}& \small{15.990}& \small{8.884}& \small{9.742} & \small{B1III}& \small{}\\
\small{J20382889+4009566} & \small{20 38 28.88} & \small{+40 09 56.63}& \small{\textit{c}}& \small{9.996}& \small{8.816} & \small{8.896}& \small{B1III}& \small{} \\
\small{J20440752+4107342} & \small{20 44 07.51} & \small{+41 07 34.18}& \small{\textit{c}}& \small{15.470}& \small{8.703} & \small{9.562}& \small{B1III}& \small{}\\
\small{LSII+3797} & \small{20 33 35.52} & \small{+38 01 36.73}& \small{\textit{c}}& \small{11.838}& \small{5.900}& \small{6.691} & \small{B1Ia}& \small{yes*}\\
\small{J20211924+3936230} & \small{20 21 19.24} & \small{+39 36 22.98}& \small{\textit{c}}& \small{12.732}& \small{7.886}& \small{8.488} & \small{B1Ib}& \small{yes*}\\      
\small{BD+394179} & \small{20 25 27.28} & \small{+40 24 00.15}& \small{\textit{c}}& \small{11.066}& \small{6.667}& \small{7.210} & \small{B1Ib}& \small{}\\
\small{J20290247+4231159} & \small{20 29 02.46} & \small{+42 31 15.91}& \small{\textit{c}}& \small{13.330}& \small{8.426}& \small{9.090} & \small{B1Ib}& \small{yes*}\\ 
\small{J20381289+4057169} & \small{20 38 12.88} & \small{+40 57 16.86}& \small{\textit{c}}& \small{12.870}& \small{8.675}& \small{9.183} & \small{B2V}& \small{}\\

\hline
		\end{tabular}		
		\tablefoot{$B$ magnitudes from the USNO-B catalog. $Ks$ and $J$ magnitudes from 2MASS catalog. Region \textit{a} indicates the 1 deg. circular area centered on Cyg OB2$\#$8 trapezium adopted by  \cite{wright15} for the Cygnus OB2 association. Region \textit{b} indicates the 1 deg. radius area adopted by \cite{com12} for the same Cygnus OB2 association. Region \textit{c} indicates the surrounding area outside the 1 deg. radius adopted by \cite{com12}. For binary stars asterisks indicate possible SB2 stars, whose spectral types are refered to the primary component. }
\end{table*}

The accuracy of the spectral classification depends on the effects of spectral resolution as well as the signal-to-noise ratio $(S/N)$.
The main diagnostic method for O-type stars is the comparison of He\,II 4542/He\,I 4471 ratio \citep{sota11}. These lines are similar for a O7 type star. For later O types the relative strengths of He\,II 4542/He\,I 4387 and He\,II 4200/He\,I 4144 are normally used, and represent the main criteria for types O8-B0. Spectral types earlier than O8 were classified using the criteria described by \cite{gray09}. The presence of metal lines in different ionization stage, such as Si\,III or Mg\,II, indicates early B-type stars. The relative strength of HeII 4471/MgII 4481 is a useful indicator for B1-late B stars. As secondary indicators we used the criteria described by \cite{gray09}, which were also used to classify stars of spectral types A, F and G.

Regarding the luminosity class, the criteria used for early O-type stars were introduced by \cite{wal71,wal73}, taking into account the emission effects in the He\,II 4686 line and N\,III 4634-4640-4642. For late O-type stars, we have used the criteria described by \cite{sota11} and for B-type stars the criteria described by \cite{gray09} and the  Balmer lines width.

 Although the classification for O-type stars was based on the described He and metal lines diagnostic criteria, we have also used for O-type stars the Marxist Ghost Buster (MGB) code developed by \cite{maiz12} in order to obtain a more accurate result. This tool compares the observed spectra with a grid of O standards (in this work the GOSSS library), allowing us to vary spectral type, luminosity class, velocity and resolution until obtaining a best match. Furthermore, for B-type stars we have used a sample of IACOB standards \citep{ssimon15} to improve their classification.

  \begin{table}[t!]
	\centering
	\caption{Candidates classified as late B and A-F-G type stars.}
	\label{table3}	
		\begin{tabular}{lcc}
		\hline   
		\hline \\[-1.5ex]  
    	\small{Object}&\small{RA (hhmmss)}& \small{Dec ($^{\circ}$ $^\prime$ $^{\prime\prime}$)} \\
    \hline  \\[-1.8ex] 
\small{\textit{  late B-type stars}} & \small{} & \small{}  \\  
      \cline{1-1}\\[-1.5ex]
\small{CCDMJ20323+4152AB} & \small{20 32 20.81} & \small{+41 52 00.78}\\
\small{BD+423785a} & \small{20 34 15.39} & \small{+43 09 35.28}\\
 \cline{1-1}   \\[-1.8ex]   
\small{\textit{  A-type stars}} & \small{} & \small{}  \\  
\cline{1-1}\\[-1.5ex]
\small{J20252497+3934030} & \small{20 25 24.96} & \small{+39 34 03.02} \\
\small{J20285874+4013302} & \small{20 28 58.74} & \small{+40 13 30.22}\\
\small{J20315984+4120354} & \small{20 31 59.84} & \small{+41 20 35.41}\\
\small{J20320734+3828586} & \small{20 32 07.34} & \small{+38 28 58.62}\\
\small{BD+413801} & \small{20 33 30.39} & \small{+42 04 17.35} \\
\small{J20364336+3906145} & \small{20 36 43.36} & \small{+39 06 14.53} \\
\small{CCDM J20429+4311AB} & \small{20 42 54.24} & \small{+43 10 38.71}\\
 \cline{1-1}  \\[-1.8ex]
\small{\textit{  F-type stars}} & \small{} & \small{}  \\  
 \cline{1-1}\\[-1.5ex]
\small{J20193232+4042447} & \small{20 19 32.32} & \small{+40 42 44.72}\\ 
\small{J20253116+4005508} & \small{20 25 31.16} & \small{+40 05 50.82}\\
\small{J20275204+4131200} & \small{20 27 52.03} & \small{+41 31 19.98}\\
\small{CCDM J20420+4015} & \small{20 42 01.18} & \small{+40 14 42.70} \\
\small{J20500396+4300118} & \small{20 50 03.96} & \small{+43 00 11.76} \\
\small{J20504551+421012.6} & \small{20 50 45.51} & \small{+42 10 12.64}\\
 \cline{1-1} \\[-1.8ex]
\small{\textit{  G-type stars}} & \small{} & \small{} \\  
 \cline{1-1}\\[-1.5ex]
\small{J20300022+4337553} & \small{20 30 00.21} & \small{+43 37 55.29}\\
\small{J20340430+4136507} & \small{20 34 04.29} & \small{+41 36 50.67}\\
\small{J20432737+4308525} & \small{20 43 27.37} & \small{+43 08 52.47}\\
\small{J20472235+4220523} & \small{20 47 22.34} & \small{+42 20 52.30}\\
\hline
		\end{tabular}		
\end{table}

\section{New confirmed OB type stars}\label{sect4}

The observed spectra have high enough $S/N$ and resolution for a spectral classification of the stars. All the candidate spectra are plotted in Appendix~\ref{appb1} (Fig.~\ref{figb1}) where the main diagnostic lines used are also indicated.

 Out of the 61 candidates 42 are OB type stars, earlier than B3, including 11 O-type stars. Two more are late B-type stars. The location of these new confirmed OB stars is shown in Fig.~\ref{fig1} while their names, coordinates, magnitudes and the derived spectral classification are listed in Table~\ref{table2}. We have also included the region in which they are located: (\textit{a}) the Cygnus OB2 area considered by \cite{wright15} which is the youngest core of Cyg OB2 at present; (\textit{b}) the Cygnus OB2 area considered by \cite{com12} which is the extended Cyg OB2 area containing older stars, on average, and (\textit{c}) the surrounding area which includes part of the Cygnus OB9 association and field population. 
The remaining observed stars are late B and foreground A-F-G stars, which are globally listed in Table~\ref{table3}. 
The selection criteria success rate obtained by \cite{com12} along with the  number of the confirmed OB type obtained in this work ($72\%$), support the success at identifying reddened massive OB stars with this method.
 
\begin{figure}[ht!]
\centering
\includegraphics[width=4.3cm, height=4.2cm ]{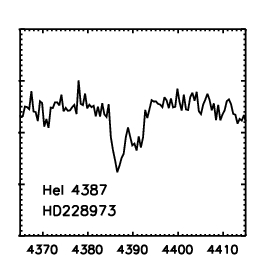}
\includegraphics[width=4.3cm, height=4.2cm ]{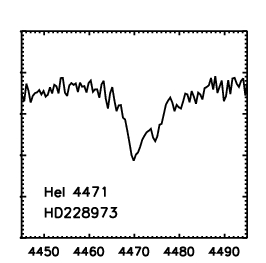}
\caption{Example of SB2 detection: HeI lines in star HD228973.}
\label{fig2}

\end{figure}

We have also detected nine possible or confirmed SB2 binaries in the sample of new OB stars (see Table~\ref{table2}). In some cases, indicated with asterisks, we can only suggest possible binary nature mainly due to noisy spectra.
This number represents a $21\%$ of our massive OB sample. We assign to these stars the spectral classification of their primary component. An example is shown in Fig.~\ref{fig2}. \cite{sana11} found that at least $45-55\%$ of the O star population in clusters and OB associations is comprized of spectroscopic binaries, which indicates that it is highly likely that more binaries are undetected in our sample. 

 \subsection{Extinction}\label{sect41}

 \begin{figure}[b!]
\centering
\includegraphics[width=8.9cm, height=7.0cm ]{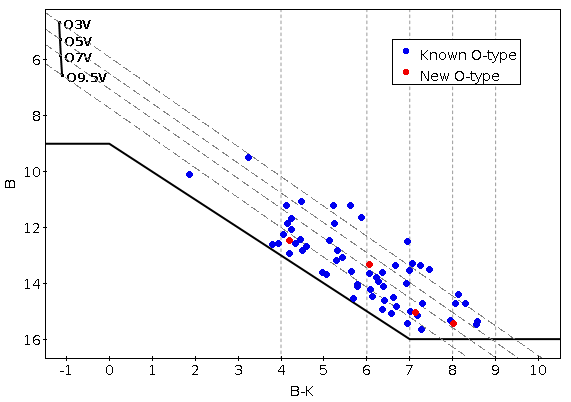}
\caption{ $(B-K)$, $B$ diagram of the confirmed O-type stars in Cygnus OB2 within the 1 deg. radius area adopted by \cite{com12}. Blue dots indicate stars previously known, and red dots represent the new ones confirmed in this work. The upper left vertical line shows the position  of the unreddened main sequence based on intrinsic magnitudes from \cite{martins06} at a distance modulus $DM = 10.8$ mag. The dash-dotted lines represent the \textit{locus} expected for different spectral type stars using the reddening law of \cite{wright15} with $R_{\rm V}$ = 2.91. The solid line marks the limits imposed by the selection criteria at magnitudes $B\leq16$ mag, $Ks<9$ mag. Vertical dotted lines indicate $(B-K)$ color bins.} 
\label{fig3} 
\end{figure}

We used the $(J-Ks)$ colors and the recent extinction law derived by \cite{wright15} ($R_{\rm V} = 2.91$) to obtain  individual extinctions for all the new confirmed massive OB type stars (see Table~\ref{tablea1}). Most of them have visual extinctions in the range $A_{\rm V}= 4-8$ mag, which agrees with previous studies in the region \citep{com12,wright15}. But, as in \cite{com12}, this is partly consequence of the imposed $B\leq16$ and $Ks<9$ magnitude limits.
 
In Fig.~\ref{fig3} we present the $B$, $(B-K)$ diagram of the confirmed O-type stars in Cygnus OB2. We observe that some of them are aligned below the O9.5V reddening vector, which could suggest a small shift in the adopted intrinsic color calibration since most of them are already known dwarf late O-type members. In spite of this we can assess the incompleteness of the O population in the region by taking into account the ratios of stars in different spectral (O3-O5, O5-O7 and O7-O9.5) and $(B-K)$ color bins.

 All the stars of our sample with $(B-K)$ colors $<6$ mag should be bright enough to have been detected by our selection criteria. Figure~\ref{fig3} shows that the population is expected to be complete for $(B-K)<7$ mag. But for $(B-K)>7$ mag we can not see late O-type stars because they are fainter than $B = 16$ mag, which is the imposed magnitude limit. Assuming that the ratios of spectral types are independent of extinction, we can conclude we are loosing the fainter or more obscured O-type stars.
 
\begin{figure}[b!]
\centering
\par{
\includegraphics[width=7.8cm ]{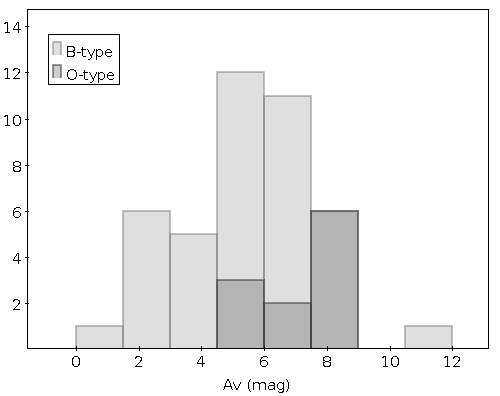}
\includegraphics[width=7.8cm ]{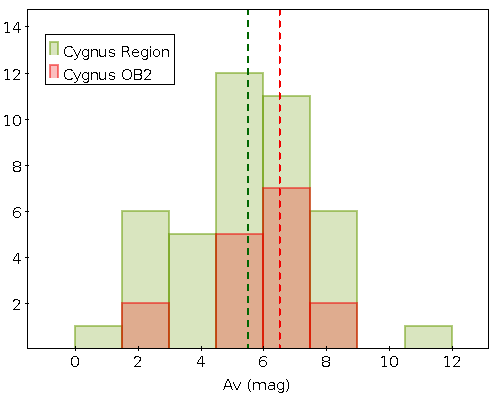}
\par}
\caption{ \textit{Top}: Extinction distribution of the new classified OB type stars earlier than B3 in Cygnus OB2 and boundaries. Gray indicates B-type stars while dark gray are O-type stars. \textit{Bottom}: Green indicates all the confirmed stars in the Cygnus region, which includes Cyg OB2, Cyg OB9 and field. Red indicates only those new OB stars located in Cygnus OB2. Green and red dashed lines indicate the median values for the stars in the whole Cygnus region and only Cygnus OB2 respectively.} 
\label{fig4} 
\end{figure}

The extinction distribution for the 42 new confirmed OB stars earlier than B3 in Cygnus OB2 and boundaries is shown in Fig.~\ref{fig4} (\textit{top}). Again, the spectral classification for those stars classified as possible or confirmed SB2 binaries is taken from the primary component. 
Although O-type stars are clearly more obscured, a median value of $A_{\rm V} = 5.5$ mag was found for the whole sample. However, this value is also affected by completeness. O-type stars are more obscured on average because we can detect them up to higher foreground extinctions thanks to their intrinsic brightness.
In the bottom histogram the same sample is differentiated by location: the whole sample that belongs to Cygnus OB2, Cygnus OB9 and boundaries, and as a sub-sample, those stars located within the Cygnus OB2 area  \citep[from][]{com12} for which a median value of $A_{\rm V} = 6.5$ mag is derived.

We see an abrupt cut for $A_{\rm V}\geq9$ mag. Stars with such high visual extinctions shall be intrinsically extremely bright to  be seen. An O9V star with a visual extinction of $A_{\rm V} = 9$ mag will have an apparent magnitude $B\simeq17$ mag, beyond the selection criteria limits and probably this is the reason why we do not find stars beyond this value. However, for the star J20395358+4222506 classified as B0I, we have obtained a visual extinction of 11.0 mag. It has a magnitude in the B band of 15.89 mag but in the Ks band of only 5.82 mag which indicates a very bright star.
These cases show again that the sample does not provide a complete census. The magnitude-limited sample of Cygnus OB2 is made incomplete due to extinction.
 
\begin{figure}[t!]
\centering
\includegraphics[width=9.5cm, height=8.0cm ]{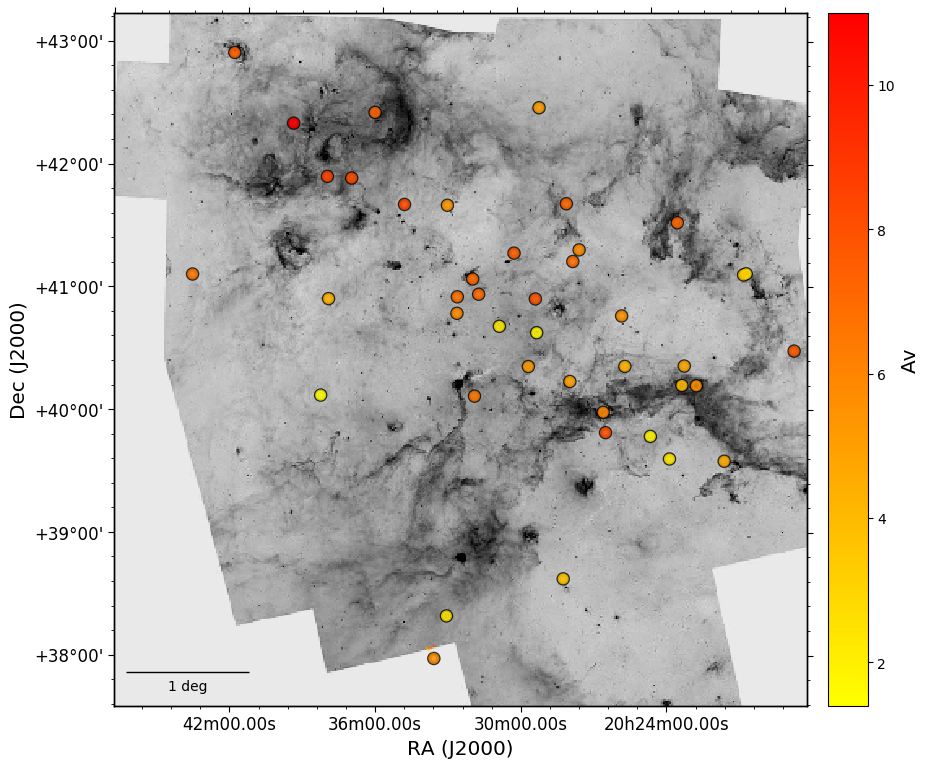}
\caption{ Location of the new confirmed OB stars in the region colored according to the derived extinction ($A_{\rm V}$) and over-plotted on an inverse Spitzer 8 $\mu$m image.}
\label{map}  
\end{figure}

In Fig.~\ref{map} we present the spatial location of the new classified OB stars, where each star is color-coded according to its derived visual extinction $A_{\rm V}$. The extinction distribution varies smoothly across the region, increasing from the south-west (Cygnus OB9) to the northeast (Cygnus X-North) where the most extinguished star of our sample, J20395358+4222506, is located.

\subsection{Stellar parameters}\label{sect43}
\subsubsection{Effective temperature}\label{sect431}

  For each classified O-type star and luminosity classes V, III and I, we have used the effective temperature ($T_{\rm eff}$) scale developed by  Holgado et al. (\textit{in prep.}) based on a revision of the O-type standards in the IACOB database. We have preferred it over the SpT-$T_{\rm eff}$ calibration suggested by \cite{martins05} since the latter gives too low $T_{\rm eff}$ values for late O-type stars \citep{ssimon14}.
  
\begin{figure}[h!]
\centering
\includegraphics[width=8.4cm, height=6.5cm ]{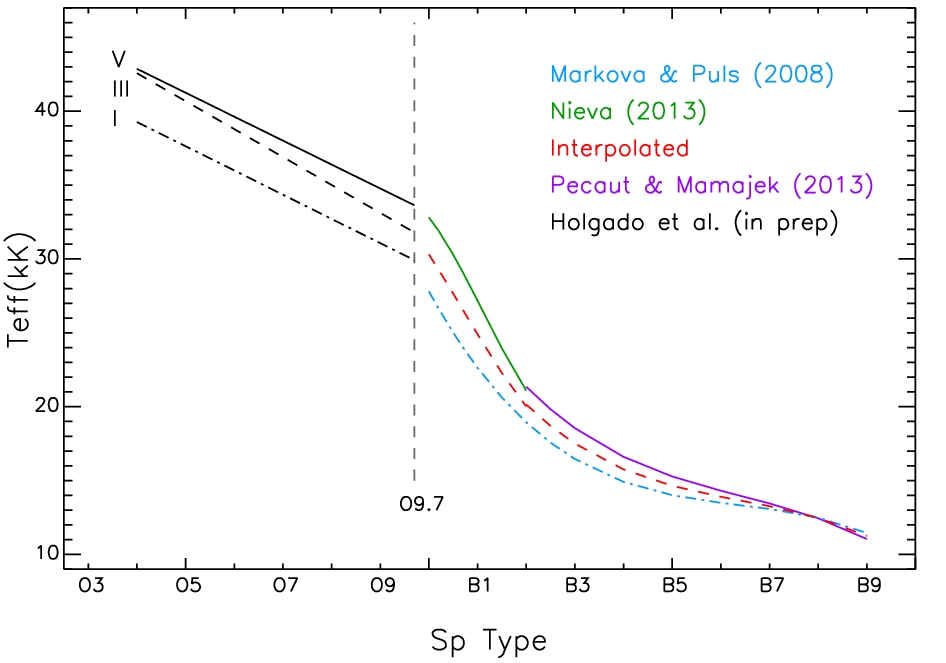} 
\caption{$T_{\rm eff}$ scales used in this work.}
\label{fig6}
\end{figure}

   For early B-type stars and luminosity class V, we decided to use the $T_{\rm eff}$ spectral type compilation of \cite{nieva13}. An excellent agreement with the calibration of Holgado et al. is found. For late B-type dwarf stars we have used the calibration of \cite{pecaut13}. Regarding B stars with luminosity class I, we have used the $T_{\rm eff}$ scale of \cite{markova08}, and for luminosity class III we have interpolated between classes I and V.  In Fig.~\ref{fig6} are represented the different $T_{\rm eff}$ scales adopted in this work, and the derived $T_{\rm eff}$ values for all the new OB type stars are shown in Appendix~\ref{appa} (Table~\ref{tablea1}). We see that all scales for B-types fit smoothly the Holgado et al. scale. Temperatures were also obtained for those stars classified as SB2 binaries by taking into account the primary component.
   
 \cite{com12} used the `observational' effective temperature based on the  $T_{\rm eff}$ versus spectral type calibration of \cite{martins05}. For B-type stars, they used the $T_{\rm eff}$ spectral type compilation of \cite{tokunaga00} but applying a scaling factor to force the agreement with \cite{martins05} temperature scale. In order to be consistent with them we have recalculated temperatures for all of their known OB stars sample in the Cygnus Region using our criteria. New values are shown in Appendix~\ref{appa} (Table~\ref{tablea2}).

\subsubsection{Bolometric correction and luminosity}\label{sect432}

As \cite{com12}, we have adopted a distance modulus of $DM = 10.8$ mag, and used the intrinsic magnitudes derived by \cite{martins06} for O stars and those compiled by \cite{tokunaga00} for B stars. Thus we could derive absolute magnitudes \citep[see][]{com08}. To derive luminosities we adopted bolometric corrections (BCs) from \cite{lanz03,lanz07}. There is an excellent agreement with \cite{martins05} for temperatures higher than $\sim$32000 K, and also with the bolometric corrections from \cite{nieva13} for cooler stars \citep[see][]{nieva13}. The derived luminosity values for the new and already known OB stars are shown in Table~\ref{tablea1} and Table~\ref{tablea2} respectively.

\begin{figure*}[ht!]
\centering
\par{ 
\includegraphics[width=8.9cm, height= 6.7cm]{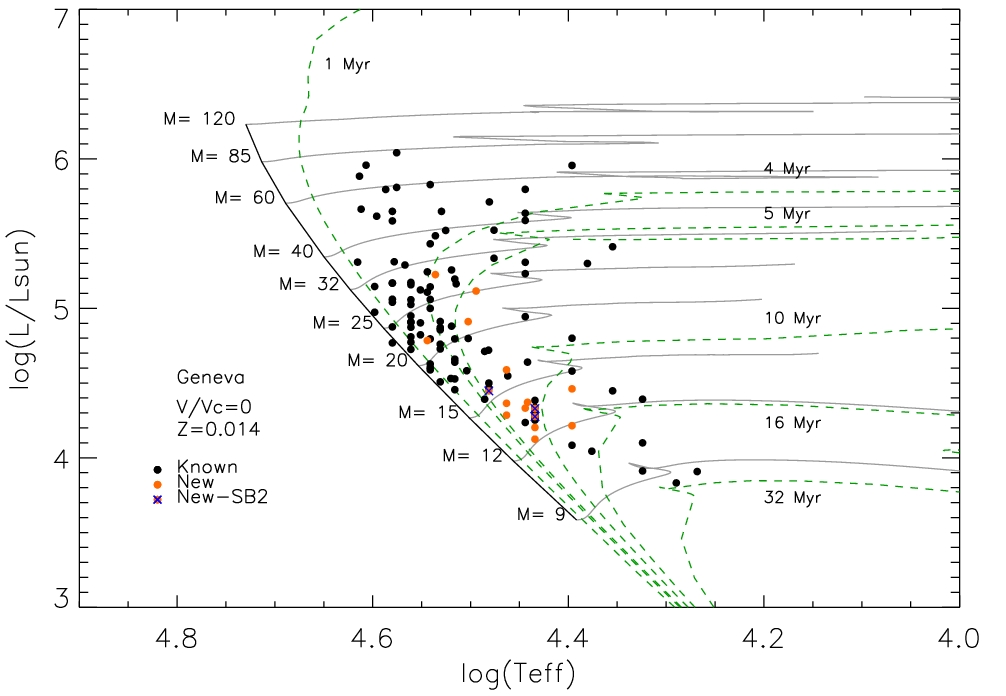}
\includegraphics[width=8.9cm, height= 6.7cm ]{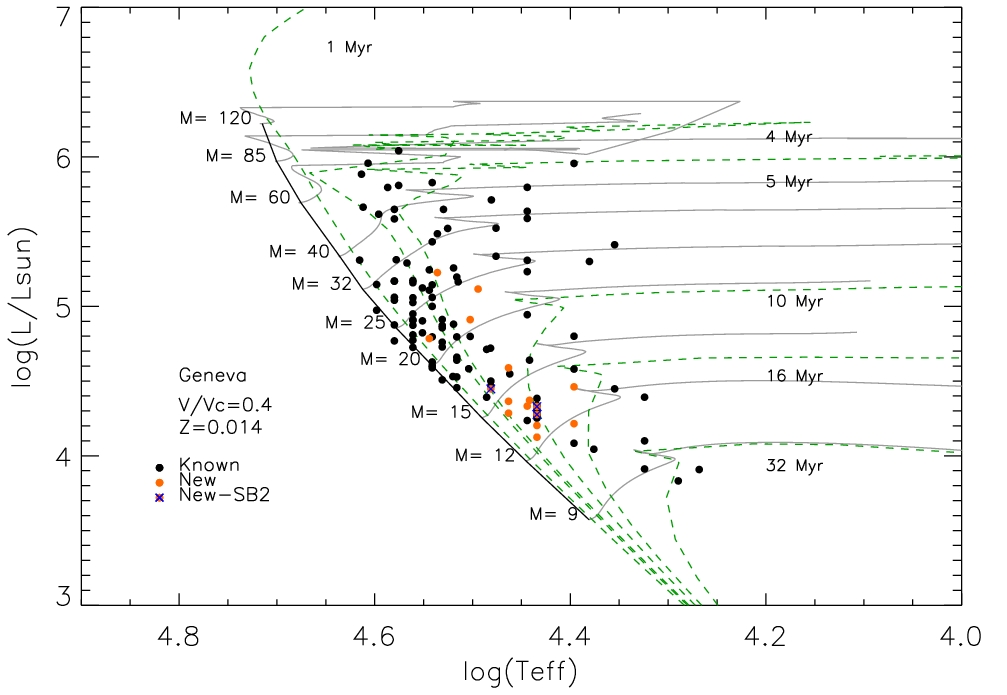} 
\includegraphics[width=8.9cm, height= 6.7cm]{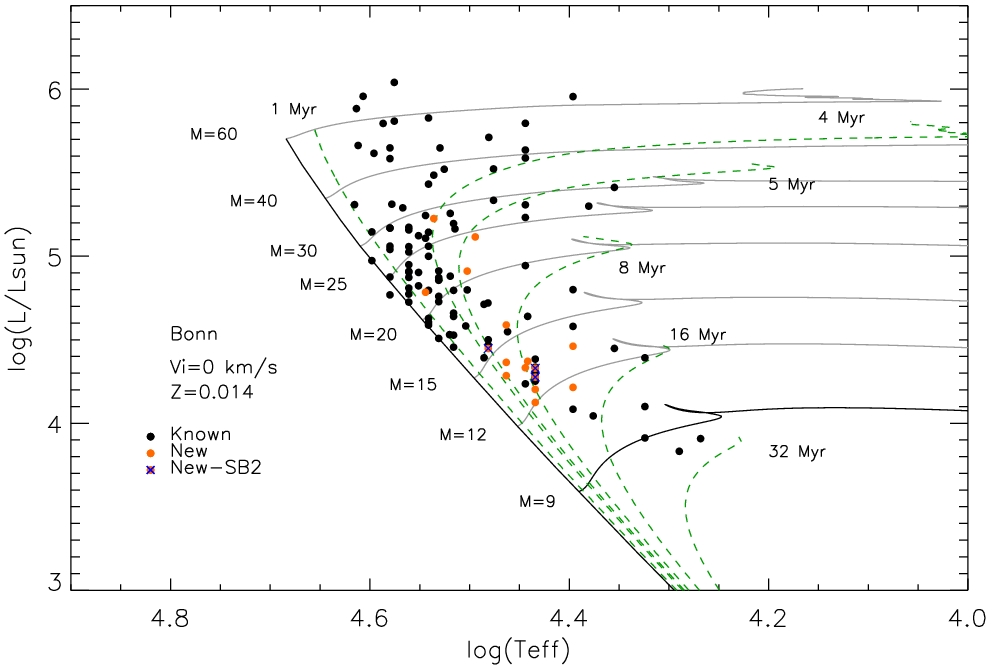}
\includegraphics[width=8.9cm, height= 6.7cm ]{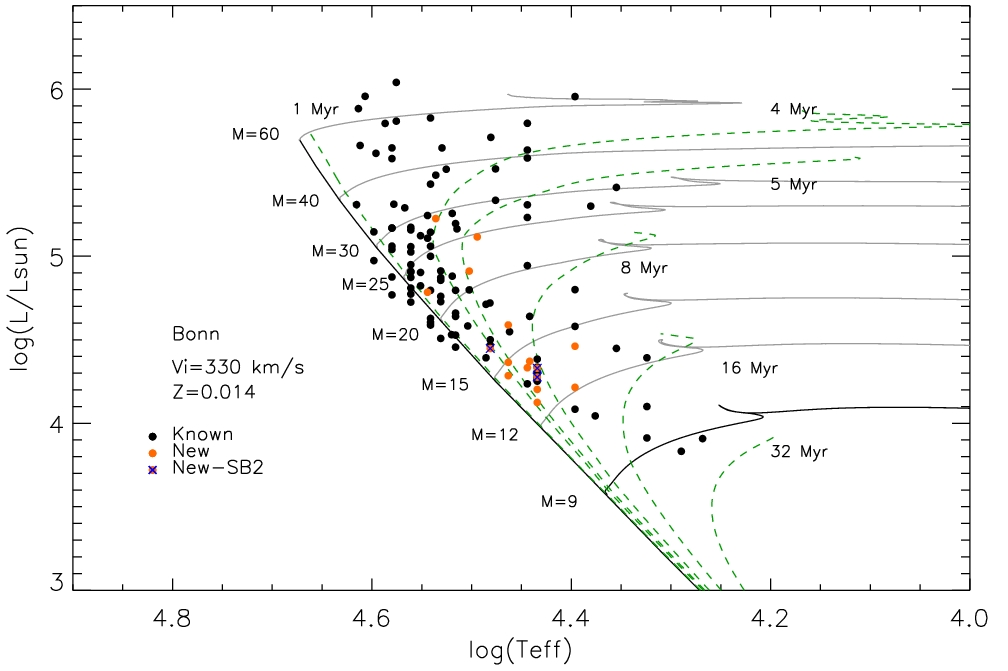}  
\par}  
\caption{HR diagrams of OB stars in Cygnus OB2 assuming a $DM = 10.8$ mag.  Known OB stars from \cite{com12} are also included to complete the sample. Black and orange dots indicate the already known and new OB-type stars respectively. Blue crosses indicate those new stars classified as possible SB2 stars. \textit{Top left hand panel}: Isochrones (dotted lines) and evolutionary stellar tracks (solid lines) for non-rotating models from \cite{ekstrom12}. \textit{Top right hand panel}: Isochrones (dotted lines) and evolutionary stellar tracks (solid lines) for rotating models (V/V$_{\rm c}$ = 0.4) from \cite{ekstrom12} .
 \textit{Bottom left hand panel}:  Isochrones (dotted lines) and evolutionary stellar tracks (solid lines) for non-rotating models from \cite{brott11}. \textit{Bottom right hand panel}:  Isochrones (dotted lines) and evolutionary stellar tracks (solid lines) for rotating models (V$_{\rm i}$ = 330 km/s) from \cite{brott11}.}
 \label{fig7}
  \end{figure*}
 
\section{Discussion}\label{sect5}
 
Some of the new confirmed OB type stars (see Fig.~\ref{fig1}) are located at the boundaries assigned to Cygnus OB2. However, the limits of the association are not strongly defined. The first surveys in the association  assumed a smaller area where the most luminous members are located \citep{mm53,mt91}. Then, \cite{knodlseder00} provided wider limits showing that extinction was a limiting factor in previous studies. This area was thus extended with the identification of new early-type members \citep{com02,com08,com12,wright09,wright15}.
If we assume the extension of Cygnus OB2 as the area  adopted by \cite{com12} (1 deg. radius centered on Galactic coordinates $l=$79.8$^{\circ}$ and $b=$+0.8$^{\circ}$), as well as the most recent census in the area developed by \cite{wright15}, we can now update the census of O and B-type stars in Cygnus OB2 from 204 to 221, and the number of confirmed O-type members increases from 66 to 70 stars.

\subsection{Hertzsprung-Russell diagram}\label{sect51}

Some studies about the age of Cygnus OB2 have been developed in the last two decades \citep{massey01, hanson03, drew08, negueruela08, wright10, com12, com16}. However, different evolutionary stellar models were used. We decided to explore four different stellar models (two families with two initial rotational velocities) to construct the Hertzsprung-Russell Diagrams (HRD) in order to assess uncertainties on the age distribution.
    
We have placed the new confirmed OB stars in the HRD to study the evolutionary status of the association. For this aim we have considered the stars within the area adopted by \cite{com12}. We have also added the already known OB stars compiled in their work to complete the sample. Two of them, classified as B0 stars (J20325964+4115146 and J20333700+4116113) were recently reclassified as O9.7III and O9.5IV by \cite{maiz16}. We thus decided to assume this last classification for both.
We have used the Geneva evolutionary tracks and isochrones with and without rotation as calculated by \cite{ekstrom12} (Fig.~\ref{fig7} \textit{top}), and for comparison, the Bonn evolutionary tracks and isochrones with and without rotation as calculated by \cite{brott11} (Fig.~\ref{fig7} \textit{bottom}).

In the Geneva case, we found a difference up to $\sim$2 Myr in the stellar ages for those most massive members ($\geq$ 20 M$_{\odot}$) depending on whether we consider rotation or not. Larger stellar lifetimes are derived from rotating stellar models. A similar result was found by \cite{wright15} by comparing both non-rotating and rotating Geneva stellar models. They derived an age range of 1 - 7 Myr for the association, but the results from rotating stellar models suggested an age of 4 - 5 Myr, while non-rotating models hinted at a younger age ($\sim$2 - 3 Myr).
On the contrary, the Bonn stellar models included on this work do not exhibit a large difference in the estimated age of the stars when they include or not stellar rotation. This different behavior is most likely to be ascribed to their different treatment of rotation. Thus, the positions of the isochrones in both Bonn HR diagrams are very similar. 
Moreover, in the Bonn case we see an extended terminal age main sequence (TAMS), and therefore, all the OB stars are located on the main sequence (MS).
In spite of this, all models suggest that most of the stars are in the age range of \mbox{1 - 6 Myr}, indicating a continuous (but not necessarily constant) star formation activity. 
However, a combination of different scenarios (rotating and non-rotating stellar models) would help to narrow the possible ages. Slow rotators with initial masses of \mbox{30 M$_{\odot}$} or less would give similar ages as more massive stars born with larger rotational velocities.

In view of the results, we can conclude that uncertainties due to rotation and adopted models affect mainly the most massive members, suggesting larger ages of about 1 - 2 Myr for them. This makes the age determination very uncertain and strengthens the need for additional diagnostics, in particular individual rotation velocity measurements, to better constrain them.

\subsection{Age distribution across Galactic longitude} \label{sect511}

\cite{com12} studied the correlation between ages and Galactic longitudes in Cygnus OB2. They suggest that star formation has proceeded from lower to higher Galactic longitudes, finding most of the old stars located at low Galactic longitudes while the youngest ones lie at higher Galactic longitudes. In order to corroborate it, we have obtained ages by comparing isochrones from the HRDs, and divided the sample of OB stars in three age groups: the young group contains stars in the range of 0 - 5 Myr, the intermediate group contains stars in the range of 5 - 10 Myr, and the old one contains stars with ages $>10$ Myr.

\begin{figure}[t!]
 \centering
 \includegraphics[width=8.5cm, height= 5.4cm]{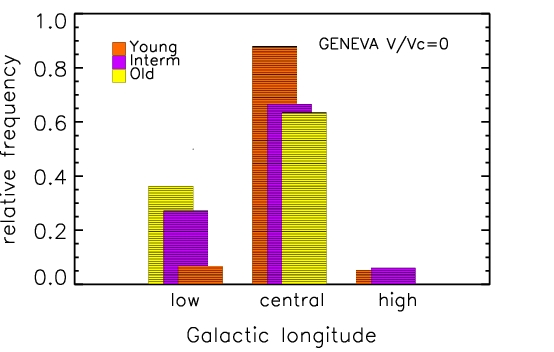}
 \includegraphics[width=8.5cm, height= 5.4cm]{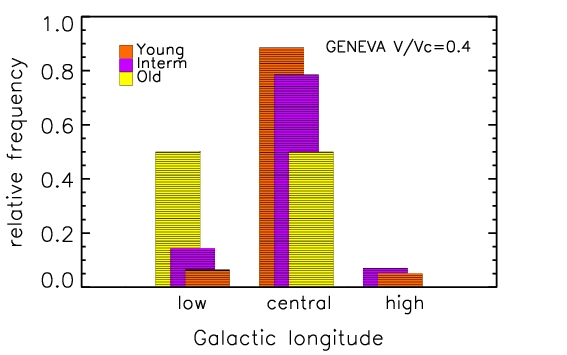} 
\caption{ Relative frequency histograms of the stars located at low (78.5$^{\circ}$- 79.5$^{\circ}$), central (79.5$^{\circ}$- 80.5$^{\circ}$) and high (80.5$^{\circ}$- 81.5$^{\circ}$) Galactic longitudes in Cygnus OB2 using Geneva non-rotating (\textit{top}) and rotating (\textit{bottom}) models. Orange, purple and yellow colors indicate those stars located in the young (0 - 5 Myr), intermediate (5 - 10 Myr) and old \mbox{(> 10 Myr)} age groups.}         
\label{fig8a}
 \end{figure}

\begin{figure}[t!]
\centering
\includegraphics[width=8.5cm, height= 5.4cm]{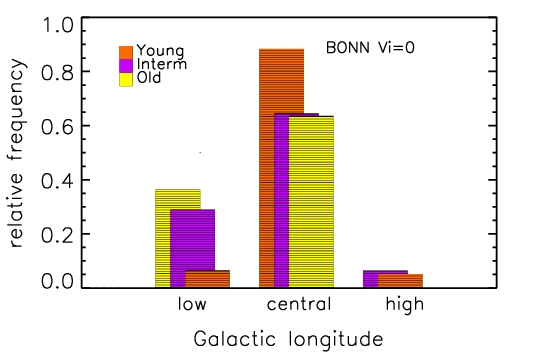}
\includegraphics[width=8.5cm, height= 5.4cm]{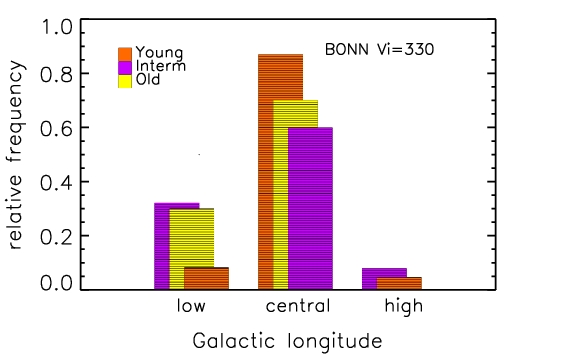} 
\caption{ Relative frequency histograms of the stars located at low (78.5$^{\circ}$- 79.5$^{\circ}$), central (79.5$^{\circ}$- 80.5$^{\circ}$) and high (80.5$^{\circ}$- 81.5$^{\circ}$) Galactic longitudes in Cygnus OB2 using Bonn non-rotating (\textit{top}) and rotating (\textit{bottom}) models. Orange, purple and yellow colors indicate those stars located in the young (0 - 5 Myr), intermediate (5 - 10 Myr) and old \mbox{(> 10 Myr)} age groups.}         
\label{fig8b}
 \end{figure}

Figures~\ref{fig8a} and~\ref{fig8b} show the histograms of the relative frequency of our OB type sample located at low (78.5$^{\circ}$- 79.5$^{\circ}$),  central (79.5$^{\circ}$- 80.5$^{\circ}$) and high (80.5$^{\circ}$- 81.5$^{\circ}$) Galactic longitudes in each age group and for each model considered (Geneva (Fig.~\ref{fig8a}) and Bonn (Fig.~\ref{fig8b}) rotating and non-rotating models). In all cases, most of the stars at high Galactic longitudes belong to the young-intermediate age group, while those stars located at low Galactic longitudes belong to the old age group. 
We only obtain significant age differences at using the rotating Geneva models since in this case rotation gives older ages for the most massive stars. Thus,  we can see a change in the relative frequencies of old and intermediate stars. On the other hand, we do not see these age differences at using rotating Bonn models because of their lower sensitivity to rotation as noted earlier.    

This analysis thus supports the correlation between ages and Galactic longitudes in Cygnus OB2 as previously suggested by \cite{com12}. Massive star formation in Cygnus OB2 seems to have proceeded from lower to higher Galactic longitudes, regardless of the details of the models used.
 
\subsection{The whole Cygnus region}\label{sect512} 

\begin{figure*}[t!]
 \centering
 \includegraphics[width=15.5cm, height= 10.5cm  ]{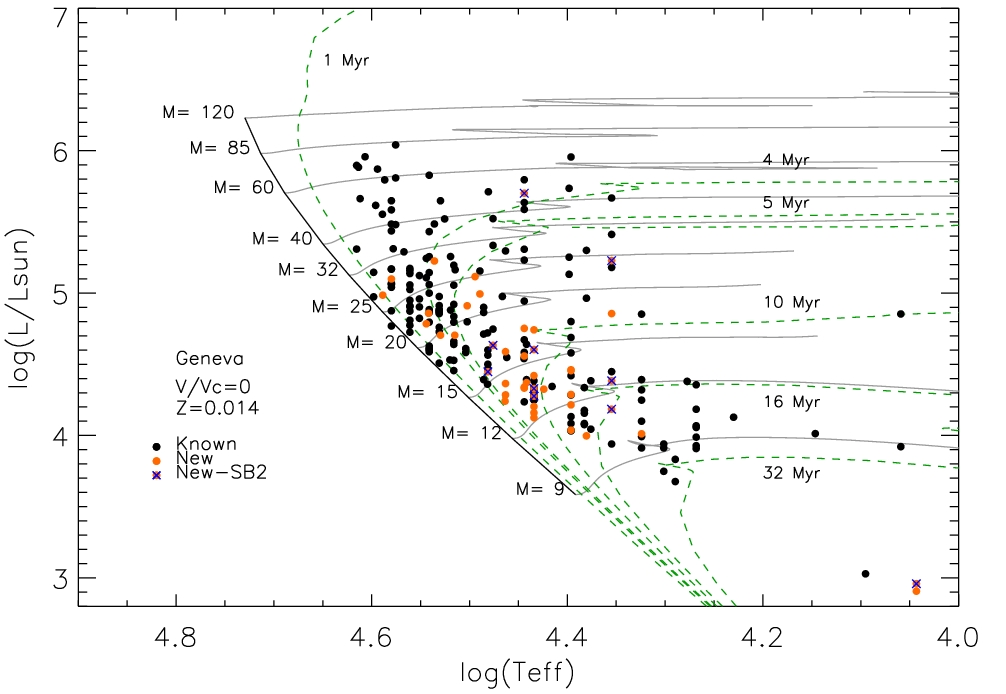} 
 \caption{ HR diagram of OB stars in the Cygnus region (Cygnus OB2, Cygnus OB9 and field population) assuming a $DM = 10.8$ mag. Known OB stars from \cite{com12} are also included to complete the sample. Black and orange dots indicate the already known and new OB-type stars respectively. Blue crosses indicate those new stars classified as possible or confirmed SB2 stars. Isochrones (dotted lines) and evolutionary stellar tracks (solid lines) for non-rotating models are from \cite{ekstrom12}. Late-B stars are also included.}
 \label{fig9}
  \end{figure*}

\begin{figure*}[t!]
 \centering
 \includegraphics[width=6.00cm, height= 4.5cm  ]{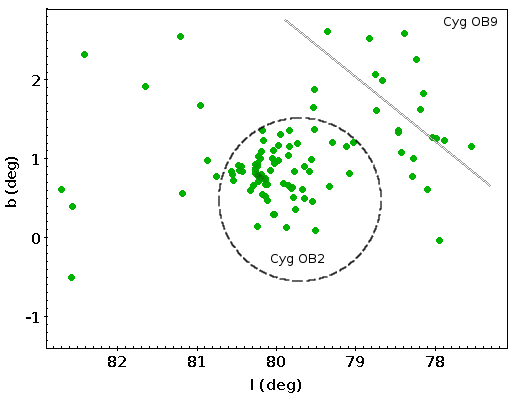} 
 \includegraphics[width=6.00cm, height= 4.5cm  ]{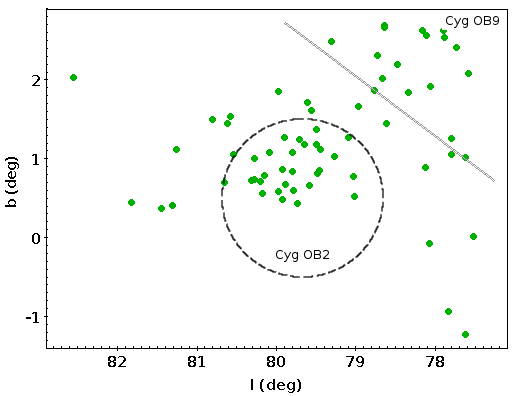} 
 \includegraphics[width=6.00cm, height= 4.5cm  ]{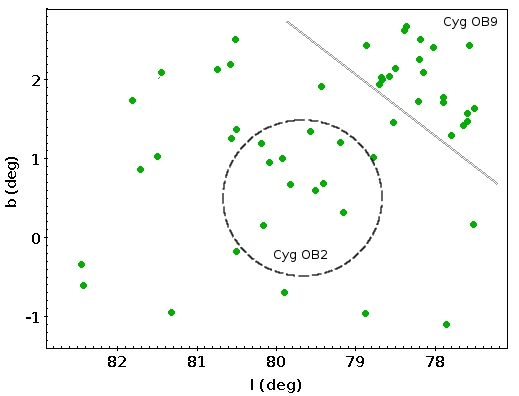} 
 \caption{ Spatial distribution of stars of different age in the Cygnus area: the young population (0 - 5 Myr) on the left, the intermediate age one \mbox{(5 - 10 Myr)} in the middle, and the old population (> 10 Myr) on the right.}
 \label{fig10}
  \end{figure*}

In order to obtain a big picture of the age distribution across the Galactic longitude, we have performed the same analysis as in Cygnus OB2 but now in a wider area which includes Cygnus OB2, Cygnus OB9 and boundaries. We have placed the whole sample of OB stars, including late B-type stars, in an HRD (see Fig.~\ref{fig9}). Due to the similar results obtained in Cygnus OB2 by using different models, we have now used isochrones and evolutionary stellar tracks for non-rotating models from \cite{ekstrom12}. We derived ages by comparison with isochrones and divided again the sample in three age groups: the young group \mbox{(0 - 5 Myr)}, the intermediate group (5 - 10 Myr), and the old group ($> 10$ Myr). In Fig.~\ref{fig10} we show the spatial distribution of the different age groups, where most of the younger stars (\textit{left}) are concentrated at higher Galactic longitudes, while the older ones (\textit{right}) are located at  central-lower Galactic longitudes. 
These results suggest that massive star formation has proceeded from lower to higher Galactic longitudes, from Cygnus OB9 to Cygnus OB2, with a strong peak in the northern part of the association.

\section{Conclusions}\label{sect6}

We have carried out several observing runs between 2013 and 2017 to obtain new blue intermediate-resolution spectra suitable for spectral classification for the magnitude-limited ($B\leq16$ mag, $Ks<9$ mag) candidate list compiled by \cite{com12}. Out of 61 candidates, we confirm 42 new massive OB-type stars, earlier than B3, including 11 new O-type members. A $21\%$ of this sample results on confirmed or possible SB2 binaries. Two other stars are late-B, seven are A-type, six are F-type, and the remaining four are G-type stars.  Therefore, the spectral classification of the homogeneously selected sample is now completed. The selection criteria success rate obtained by \cite{com12} along with this work is $72\%$, supporting the success at identifying reddened massive OB stars with the reddening-free parameters based on $BJHK$ photometry (see Sect.~\ref{sect2}).
However, the magnitude cutoff and dust extinction introduce an incompleteness. We loose the faintest and more obscured late O-type members in Cygnus OB2, and we are still far to obtain a complete census of the early-type population in the association.
 
We have also estimated individual visual extinctions ($A_{\rm V}$) for the new confirmed OB-type stars using the extinction law derived by \cite{wright15}. We found a median value of $\sim$5.5 mag for the stars in the whole Cygnus region, and as we expect, a large median value of $\sim$6.5 mag for those stars within the Cygnus OB2 association.

We have placed the new sample of OB stars in the HR diagram using both rotating and non-rotating models calculated by \cite{ekstrom12} and \cite{brott11} in order to assess uncertainties. We have also placed the already known OB type stars compiled by \cite{com12} for completeness. To this aim, we derived effective temperatures and luminosities using spectral type calibrations for the whole sample. Although coming from different sources, our adopted calibrations fit very well with each other.
Uncertainties about rotation and adopted models affect only the most massive members, suggesting larger ages for them of $\sim$1-2 Myr. Even so, all models support the previous age spread observed in Cygnus OB2 of $\sim$1 - 6 Myr. 

In order to check the correlation between age and Galactic longitude found by \cite{com12} in Cygnus OB2, we have divided the sample of OB stars in different location and age groups by comparing with isochrones: the young group (0 - 5 Myr), the intermediate group (5 - 10 Myr), and the old group ($> 10$ Myr). Despite the differences in physical process treatment (rotation and physics of the stellar interior) done by the Geneva and Bonn groups, the spatial distribution of each age group is globally similar. Therefore, the result obtained by \cite{com12} can not be an effect of using the Geneva stellar models without rotation when age dating the Cygnus region stars.
Most of the stars at high Galactic longitudes belong to the young-intermediate age group, while stars at low Galactic longitudes belong to the old group. Assuming a wider area which includes Cygnus OB2, part of Cygnus OB9 and boundaries, we obtain similar results suggesting that massive star formation in the region proceeds from Cygnus OB9 to Cygnus OB2, with a strong peak in the northern part of the association. Therefore the correlation between age and Galactic longitude is confirmed, regardless of the details of the models used. \\

\begin{acknowledgements}

We acknowledge financial support from the Spanish Ministry of Economy and Competitiveness (MINECO) under the grants AYA2012-39364-C02-01, AYA 2015-68012-C2-01 and Severo Ochoa SEV-2015-0548. A.S. also acknowledges support from MINECO through grants AYA2013-40\,611-P and AYA2016-75\,931-C2-2-P. AP and CBM acknowledge support from the Sonderforschungsbereich SFB 881 "The Milky Way System" (subproject B5) of the German Research Foundation (DFG).
We also thank the WHT and its service programme (SW2017a04), J. Maíz Apellániz for his useful MGB software, and N. Wright for helpful discussion and comments in this paper.
\end{acknowledgements}

%
%

\def\bibname{References}


\begin{thebibliography}{}



\bibitem[Brott et al.(2011)]{brott11} Brott, I., de Mink, S.~E., Cantiello, M., et al.\ 2011, \aap, 530, A115

\bibitem[Comer{\'o}n \& Pasquali(2012)]{com12} Comer{\'o}n, F., \& Pasquali, A.\ 2012, \aap, 543, A101 

\bibitem[Comer{\'o}n et al.(2002)]{com02} Comer{\'o}n, F., Pasquali, A., Rodighiero, G., et al.\ 2002, \aap, 389, 874

\bibitem[Comer{\'o}n et al.(2008)]{com08} Comer{\'o}n, F., Pasquali, A., Figueras, F., \& Torra, J.\ 2008, \aap, 486, 453

\bibitem[Comer{\'o}n et al.(2016)]{com16} Comer{\'o}n, F., Djupvik, A.~A., Schneider, N., \& Pasquali, A.\ 2016, \aap, 586, A46 



\bibitem[Crowther et al.(2006)]{crowther06} Crowther, P.~A., Lennon, D.~J., \& Walborn, N.~R.\ 2006, \aap, 446, 279
  
\bibitem[Drew et al.(2008)]{drew08} Drew, J.~E., Greimel, R., Irwin, M.~J., \& Sale, S.~E.\ 2008, \mnras, 386, 1761     

\bibitem[Ekstr{\"o}m et al.(2012)]{ekstrom12} Ekstr{\"o}m, S., Georgy, C., Eggenberger, P., et al.\ 2012, \aap, 537, A146

\bibitem[Gray \& Corbally(2009)]{gray09} Gray, R.~O., \& Corbally, C., J.\ 2009, Stellar Spectral Classification.~Princeton University Press, 2009.  

\bibitem[Hanson(2003)]{hanson03} Hanson, M.~M.\ 2003, \apj, 597, 957

\bibitem[Herrero et al.(1999)]{herrero99} Herrero, A., Corral, L.~J., Villamariz, M.~R., \& Mart{\'{\i}}n, E.~L.\ 1999, \aap, 348, 542	

\bibitem[Herrero et al.(2001)]{herrero01} Herrero, A., Puls, J., Corral, L.~J., Kudritzki, R.~P., \& Villamariz, M.~R.\ 2001, \aap, 366, 623 

\bibitem[Herrero et al.(2002)]{herrero02} Herrero, A., Puls, J., \& Najarro, F.\ 2002, \aap, 396, 949 

\bibitem[Humphreys \& McElroy(1984)]{humphreys84} Humphreys, R.~M., \& McElroy, D.~B.\ 1984, \apj, 284, 565



\bibitem[Kiminki et al.(2007)]{kiminki07} Kiminki, D.~C., Kobulnicky, H.~A., Kinemuchi, K., et al.\ 2007, \apj, 664, 1102  

\bibitem[Kn{\"o}dlseder(2000)]{knodlseder00} Kn{\"o}dlseder, J.\ 2000, \aap, 360, 539 

\bibitem[Lanz \& Hubeny(2003)]{lanz03} Lanz, T., \& Hubeny, I.\ 2003, \apjs, 146, 417

\bibitem[Lanz \& Hubeny(2007)]{lanz07} Lanz, T., \& Hubeny, I.\ 2007, \apjs, 169, 83 

\bibitem[Lejeune \& Schaerer(2001)]{lejeune01} Lejeune, T., \& Schaerer, D.\ 2001, \aap, 366, 538 

\bibitem[Ma{\'{\i}}z Apell{\'a}niz et al.(2012)]{maiz12} Ma{\'{\i}}z Apell{\'a}niz, J., Pellerin, A., Barb{\'a}, R.~H., et al.\ 2012, Proceedings of a Scientific Meeting in Honor of Anthony F.~J.~Moffat, 465, 484 

\bibitem[Ma{\'{\i}}z Apell{\'a}niz et al.(2016)]{maiz16} Ma{\'{\i}}z Apell{\'a}niz, J., Sota, A., Arias, J.~I., et al.\ 2016, \apjs, 224, 4

\bibitem[Markova \& Puls(2008)]{markova08} Markova, N., \& Puls, J.\ 2008, \aap, 478, 823	 

\bibitem[Martins \& Plez(2006)]{martins06} Martins, F., \& Plez, B.\ 2006, \aap, 457, 637 

\bibitem[Martins et al.(2005)]{martins05} Martins, F., Schaerer, D., \& Hillier, D.~J.\ 2005, \aap, 436, 1049 

\bibitem[Massey \& Thompson(1991)]{mt91} Massey, P., \& Thompson, A.~B.\ 1991, \aj, 101, 1408 

\bibitem[Massey et al.(1995)]{massey95} Massey, P., Johnson, K.~E., \& Degioia-Eastwood, K.\ 1995, \apj, 454, 151	

\bibitem[Massey et al.(2001)]{massey01} Massey, P., DeGioia-Eastwood, K., \& Waterhouse, E.\ 2001, \aj, 121, 1050  

 
  
\bibitem[Morgan et al.(1954)]{morgan54} Morgan, W.~W., Johnson, H.~L., \& Roman, N.~G.\ 1954, \pasp, 66, 85 

\bibitem[M{\"u}nch \& Morgan(1953)]{mm53} M{\"u}nch, L., \& Morgan, W.~W.\ 1953, \apj, 118, 161 

\bibitem[Negueruela et al.(2008)]{negueruela08} Negueruela, I., Marco, A., Herrero, A., \& Clark, J.~S.\ 2008, \aap, 487, 575 	

\bibitem[Nieva(2013)]{nieva13} Nieva, M.-F.\ 2013, \aap, 550, A26 	 
  
\bibitem[Pecaut \& Mamajek(2013)]{pecaut13} Pecaut, M.~J., \& Mamajek, E.~E.\ 2013, \apjs, 208, 9



\bibitem[Portegies Zwart et al.(2010)]{port10} Portegies Zwart, S.~F., McMillan, S.~L.~W., \& Gieles, M.\ 2010, \araa, 48, 431 

\bibitem[Reddish(1968)]{reddish68} Reddish, V.~C.\ 1968, The Observatory, 88, 139



\bibitem[Reipurth \& Schneider(2008)]{reipurth08} Reipurth, B., \& Schneider, N.\ 2008, Handbook of Star Forming Regions, Volume I, 4, 36

\bibitem[Rieke \& Lebofsky(1985)]{rieke85} Rieke, G.~H., \& Lebofsky, M.~J.\ 1985, \apj, 288, 618 

\bibitem[Rygl et al.(2012)]{rygl12} Rygl, K.~L.~J., Brunthaler, A., Sanna, A., et al.\ 2012, \aap, 539, A79 

\bibitem[Sana \& Evans(2011)]{sana11} Sana, H., \& Evans, C.~J.\ 2011, Active OB Stars: Structure, Evolution, Mass Loss, and Critical Limits, 272, 474 

\bibitem[Schneider et al.(2006)]{schneider06} Schneider, N., Bontemps, S., Simon, R., et al.\ 2006, \aap, 458, 855 

\bibitem[Schulte(1956)]{schulte56} Schulte, D.~H.\ 1956, \apj, 124, 530

\bibitem[Schulte(1958)]{schulte58} Schulte, D.~H.\ 1958, \apj, 128, 41 

 \bibitem[Sim{\'o}n-D{\'{\i}}az \& Herrero(2007)]{ssimon07} Sim{\'o}n-D{\'{\i}}az, S., \& Herrero, A.\ 2007, \aap, 468, 1063 
 
 \bibitem[Sim{\'o}n-D{\'{\i}}az et al.(2014)]{ssimon14} Sim{\'o}n-D{\'{\i}}az, S., Herrero, A., Sab{\'{\i}}n-Sanjuli{\'a}n, C., et al.\ 2014, \aap, 570, L6 
 
 \bibitem[Sim{\'o}n-D{\'{\i}}az et al.(2015)]{ssimon15} Sim{\'o}n-D{\'{\i}}az, S., Negueruela, I., Ma{\'{\i}}z Apell{\'a}niz, J., et al.\ 2015, Highlights of Spanish Astrophysics VIII, 576    

\bibitem[Skrutskie et al.(2006)]{2MASS} Skrutskie, M.~F., Cutri, R.~M., Stiening, R., et al.\ 2006, \aj, 131, 1163
 
\bibitem[Sota et al.(2011)]{sota11} Sota, A., Ma{\'{\i}}z Apell{\'a}niz, J., Walborn, N.~R., et al.\ 2011, \apjs, 193, 24

\bibitem[Tokunaga(2000)]{tokunaga00} Tokunaga, A.~T.\ 2000, Allen's Astrophysical Quantities, 143 

\bibitem[Walborn(1971)]{wal71} Walborn, N.~R.\ 1971, \apjs, 23, 257

\bibitem[Walborn(1973)]{wal73} Walborn, N.~R.\ 1973, \aj, 78, 1067 

\bibitem[Walborn et al.(2002)]{wal02} Walborn, N.~R., Howarth, I.~D., Lennon, D.~J., et al.\ 2002, \aj, 123, 2754

\bibitem[Wright et al.(2009)]{wright09} Wright, N.~J., Drake, J.~J., \& Drew, J.~E.\ 2009, American Astronomical Society Meeting Abstracts \#213, 213, 605.10

\bibitem[Wright et al.(2010)]{wright10} Wright, N.~J., Drake, J.~J., Drew, J.~E., \& Vink, J.~S.\ 2010, \apj, 713, 871

\bibitem[Wright et al.(2015)]{wright15} Wright, N.~J., Drew, J.~E., \& Mohr-Smith, M.\ 2015, \mnras, 449, 741

 
 
\bibitem[Zacharias et al.(2004)]{USNO10} Zacharias, N., Urban, S.~E., Zacharias, M.~I., et al.\ 2004, \aj, 127, 3043
 
\bibitem[Zacharias et al.(2010)]{USNO04} Zacharias, N., Finch, C., Girard, T., et al.\ 2010, \aj, 139, 2184 

 
  
\end{thebibliography}

\newpage

\begin{appendix}

\section{Table of stellar parameters}\label{appa}

Table~\ref{tablea1} shows the derived stellar parameters for the new classified OB stars in a large area which includes Cygnus OB2 and its surroundings. We have also included the derived individual visual extinctions. Table~\ref{tablea2} shows the updated temperatures and luminosities for the already known OB stars from \cite{com12} assuming the same spectral calibrations used in this work (see Sect.~\ref{sect43}).

    \begin{table*}[ht!]
\centering 	
\caption{Temperatures, luminosities, and individual visual extinctions derived for the  new classified massive OB stars.}
\label{tablea1}
		\begin{tabular}{lcccc}
		\hline   
		\hline \\[-1.5ex]
    	\small{Object}& \small{SpT}& \small{$T_{\rm eff}$ (K)} & \small{log ($L/L_{\odot}$)} & \small{$A_{\rm V}$ (mag)}  \\
    \hline  \\[-1.8ex]  
\small{\textit{  Cygnus OB2}} & \small{} & \small{}  & \small{} \\  
    \cline{1-1} \\[-1.5ex] 
    
\small{J20345785+4143543}& \small{O7:Ib}  & \small{34990} & \small{5.25}& \small{8.2}\\  
\small{J20293563+4024315}& \small{O8IIIz} & \small{33961} & \small{4.75}& \small{5.1}\\ 
\small{J20275292+4144067}& \small{O9.5II} & \small{30626} & \small{5.09}& \small{7.1} \\     
\small{J20291617+4057372}& \small{O9.7III}  & \small{28518} & \small{4.78}& \small{7.7}  \\
\small{J20273787+4115468}& \small{B0II} & \small{25932} & \small{4.62}& \small{6.8}    \\     
\small{J20301097+4120088}& \small{B0:II:} & \small{25932} & \small{4.39}& \small{7.4}  \\  
\small{J20323968+4050418}& \small{B0II}  & \small{25932} & \small{4.32}& \small{5.7}  \\      
\small{J20323882+4058469}& \small{B0Ib} & \small{27800} & \small{4.33}& \small{6.7} \\ 
\small{J20272099+4121262} & \small{B0.5V} & \small{26308} & \small{4.33} & \small{5.8}  \\
\small{J20330526+4143367}& \small{B0.5III} & \small{24164} & \small{4.27}& \small{5.2}  \\     
\small{BD+404208} & \small{B1V} & \small{23590} & \small{4.14} & \small{2.2}  \\
\small{J20314341+4100021} & \small{B1V} & \small{23590} & \small{4.21}& \small{7.0}  \\
\small{J20315898+4107314} & \small{B1V} & \small{23590} & \small{4.26} & \small{7.1}  \\    
\small{BD+404193} & \small{B1V} & \small{23590} & \small{4.06} & \small{1.7}  \\ 
\small{J20274925+4017004} & \small{B1III} & \small{21557} & \small{4.27} & \small{4.9} \\
\small{J20315433+4010067}& \small{B1III} & \small{21557} & \small{4.02}& \small{6.5}   \\ 
\small{CCDMJ20323+4152AB} & \small{B9V} & \small{10700} & \small{2.96} & \small{-0.2} \\ 
  \cline{1-1}  \\[-1.8ex]  
\small{\textit{  Surroundings}} & \small{} & \small{}  & \small{} & \small{} \\  
   \cline{1-1} \\[-1.5ex] 
\small{J20423509+4256364}& \small{O6IIIz} & \small{38192} & \small{4.97}& \small{7.4}  \\ 
\small{J20371773+4156316}& \small{O7V} & \small{36872} & \small{5.06}& \small{8.2}  \\ 
\small{J20222481+4013426}& \small{O8II}  & \small{33570} & \small{4.69}& \small{5.5}   \\        
\small{J20261976+3951425}&\small{O8.5IV}& \small{33391} & \small{4.81}& \small{8.0}   \\     
\small{J20262484+4001413}& \small{O9.2III} & \small{31317} & \small{4.65}& \small{5.9}   \\
\small{J20382173+4157069}& \small{O9.7II} & \small{30859} & \small{4.99}& \small{8.5} \\ 
\small{J20181090+4029063}& \small{O9.7Ib} & \small{28644} & \small{4.58}& \small{7.6}   \\
\small{J20395358+4222506}& \small{B0I}  & \small{25094} & \small{5.57}& \small{11.0}\\
\small{J20281176+3840227}& \small{B0Ib} & \small{25094} & \small{4.43}& \small{3.6}   \\
\small{J20290247+4231159} & \small{B1Ib} & \small{19979} & \small{4.03} & \small{5.0}\\
\small{J20225451+4023314}& \small{B0Iab} & \small{25094} & \small{4.21}& \small{4.7}   \\    
\small{J20253320+4048444}& \small{B0Iab} & \small{25094} & \small{4.62}& \small{5.5}   \\
\small{HDE229258} & \small{B0.7V} & \small{24949} & \small{4.13} & \small{1.8}  \\
\small{J20361806+4228483}&\small{B0.7III} & \small{22860} & \small{4.14}& \small{7.2} \\
\small{J20233816+3938118} & \small{B0.7Ib} & \small{21514} & \small{3.87} & \small{2.3}  \\
\small{HD228973} & \small{B1V} & \small{23590} & \small{4.45} & \small{2.4}  \\
\small{J20201435+4107155} & \small{B1V} & \small{23590} & \small{4.27} & \small{3.4}  \\        
\small{J20230290+4133466} & \small{B1V} & \small{23590} & \small{4.68} & \small{7.2} \\
\small{J20330453+3822269} & \small{B1V} & \small{23590} & \small{4.09} & \small{2.4}\\
\small{J20230183+4014029} & \small{B1III} & \small{21557} & \small{4.26} & \small{4.3} \\ 
\small{J20382889+4009566} & \small{B1III} & \small{21557} & \small{3.84} & \small{1.4} \\
\small{J20440752+4107342} & \small{B1III} & \small{21557} & \small{4.09} & \small{6.5} \\
\small{LSII+3797} & \small{B1Ia} & \small{19979} & \small{5.07} & \small{5.8}   \\
\small{J20211924+3936230} & \small{B1Ib} & \small{19979} & \small{4.23} & \small{4.6} \\
\small{BD+394179} & \small{B1Ib} & \small{19979} & \small{4.70} & \small{4.2} \\
\small{J20381289+4057169} & \small{B2V} & \small{20549} & \small{3.98} & \small{4.2} \\
\small{BD+423785a} & \small{B9V} & \small{10700} & \small{2.90} & \small{0.6} \\

\hline
		\end{tabular}			
\end{table*}

    \begin{table*}[ht!]
   	\centering
   		\caption{Temperatures and luminosities derived for the known massive OB stars from \cite{com12}.}
   		\label{tablea2}
	\begin{tabular}{lccccc}
	\hline   
	\hline \\[-1.5ex]
   	\small{Object}& \small{RA (hhmmss)}& \small{Dec ($^{\circ}$ $^\prime$ $^{\prime\prime}$)}& \small{SpT} &\small{$T_{\rm eff}$ (K)} &\small{log ($L/L_{\odot}$)} \\
   \hline  \\[-1.8ex]  
\small{\textit{  Cygnus OB2}} & \small{} & \small{}  & \small{} & \small{}& \small{}   \\  
     \cline{1-1} \\[-1.5ex]
  \small{J20331411+4120218} & \small{20:33:14.110} & \small{+41:20:21.91} & \small{O3If        } & \small{40910} & \small{5.66} \\
  \small{J20330879+4113179} & \small{20:33:08.818} & \small{+41:13:17.93} & \small{O4III       } & \small{41070} & \small{5.88} \\
  \small{J20360451+4056129} & \small{20:36:04.500} & \small{+40:56:13.01} & \small{O5V((f))    } & \small{41250} & \small{5.31} \\
  \small{J20331798+4118311} & \small{20:33:17.982} & \small{+41:18:31.19} & \small{O5III       } & \small{39440} & \small{5.62} \\
  \small{J20340850+4136592} & \small{20:34:08.514} & \small{+41:36:59.39} & \small{O5I         } & \small{37630} & \small{5.80} \\
  \small{J20331074+4115081} & \small{20:33:10.735} & \small{+41:15:08.22} & \small{O5If        } & \small{37630} & \small{6.04} \\
  \small{J20332346+4109130} & \small{20:33:23.471} & \small{+41:09:12.90} & \small{O5.5V       } & \small{40440} & \small{5.96} \\
  \small{J20331326+4113287} & \small{20:33:13.264} & \small{+41:13:28.67} & \small{O6V         } & \small{39630} & \small{5.14} \\
  \small{J20303980+4136506} & \small{20:30:39.805} & \small{+41:36:50.63} & \small{O6V         } & \small{39630} & \small{4.97} \\
  \small{J20344410+4051584} & \small{20:34:44.146} & \small{+40:51:58.67} & \small{O6.5III(f)  } & \small{37845} & \small{5.31} \\
  \small{J20283203+4049027} & \small{20:28:32.027} & \small{+40:49:02.88} & \small{O7          } & \small{38010} & \small{5.65} \\
  \small{J20310019+4049497} & \small{20:31:00.204} & \small{+40:49:49.70} & \small{O7V((f))    } & \small{38010} & \small{5.06} \\
  \small{J20341350+4135027} & \small{20:34:13.511} & \small{+41:35:02.86} & \small{O7V         } & \small{38010} & \small{5.04} \\
  \small{J20331748+4117093} & \small{20:33:17.483} & \small{+41:17:09.35} & \small{O7V         } & \small{38010} & \small{5.17} \\
  \small{J20334086+4130189} & \small{20:33:40.863} & \small{+41:30:18.95} & \small{O7V         } & \small{38010} & \small{4.77} \\
  \small{J20342959+4131455} & \small{20:34:29.599} & \small{+41:31:45.49} & \small{O7V         } & \small{38010} & \small{5.58} \\
  \small{J20315961+4114505} & \small{20:31:59.609} & \small{+41:14:50.45} & \small{O7V         } & \small{38010} & \small{4.87} \\
  \small{J20321383+4127120} & \small{20:32:13.822} & \small{+41:27:12.01} & \small{O7IIIf      } & \small{36900} & \small{5.29} \\
  \small{J20313690+4059092} & \small{20:31:36.911} & \small{+40:59:09.06} & \small{O7Ib(f)     } & \small{34350} & \small{5.48} \\
  \small{J20323154+4114082} & \small{20:32:31.531} & \small{+41:14:08.18} & \small{O7.5Ib-II(f)} & \small{33530} & \small{5.52} \\
  \small{J20323857+4125137} & \small{20:32:38.571} & \small{+41:25:13.79} & \small{O8V(n)      } & \small{36390} & \small{5.02} \\
  \small{J20274361+4035435} & \small{20:27:43.616} & \small{+40:35:43.51} & \small{O8V:        } & \small{36390} & \small{4.72} \\
  \small{J20331369+4113057} & \small{20:33:13.688} & \small{+41:13:05.77} & \small{O8V         } & \small{36390} & \small{4.87} \\
  \small{J20324545+4125374} & \small{20:32:45.450} & \small{+41:25:37.57} & \small{O8V         } & \small{36390} & \small{5.16}
\small{}\\  
  \small{J20331803+4121366} & \small{20:33:18.035} & \small{+41:21:36.67} & \small{O8V         } & \small{36390} & \small{4.90} \\
  \small{J20323486+4056174} & \small{20:32:34.865} & \small{+40:56:17.35} & \small{O8V         } & \small{36390} & \small{4.77} \\
  \small{J20300788+4123504} & \small{20:30:07.877} & \small{+41:23:50.44} & \small{O8V         } & \small{36390} & \small{4.91} \\
  \small{J20330292+4117431} & \small{20:33:02.913} & \small{+41:17:43.16} & \small{O8V         } & \small{36390} & \small{5.06} \\
  \small{J20325002+4123446} & \small{20:32:50.016} & \small{+41:23:44.70} & \small{O8V         } & \small{36390} & \small{4.95} \\
  \small{J20325919+4124254} & \small{20:32:59.057} & \small{+41:24:24.79} & \small{O8V         } & \small{36390} & \small{4.81} \\
  \small{J20333030+4135578} & \small{20:33:30.316} & \small{+41:35:57.88} & \small{O8V         } & \small{36390} & \small{5.17} \\
  \small{J20342193+4117016} & \small{20:34:21.934} & \small{+41:17:01.66} & \small{O8III+O8III } & \small{35010} & \small{5.11} \\
  \small{J20323843+4040445} & \small{20:32:38.441} & \small{+40:40:44.48} & \small{O8III       } & \small{35010} & \small{5.24} \\
  \small{J20330292+4047254} & \small{20:33:02.928} & \small{+40:47:25.29} & \small{O8II((f))   } & \small{33860} & \small{5.65} \\
  \small{J20314540+4118267} & \small{20:31:45.403} & \small{+41:18:26.73} & \small{O8I         } & \small{32710} & \small{5.16} \\
  \small{J30332557+4133269} & \small{20:33:25.569} & \small{+41:33:26.88} & \small{O8.5V       } & \small{35580} & \small{5.12} \\
  \small{J20331634+4119017} & \small{20:33:16.256} & \small{+41:19:00.16} & \small{O8.5V       } & \small{35580} & \small{4.82} \\
  \small{J20332674+4110595} & \small{20:33:26.756} & \small{+41:10:59.42} & \small{O8.5V       } & \small{35580} & \small{4.90} \\
  \small{J20313749+4113210} & \small{20:31:37.506} & \small{+41:13:20.99} & \small{O9:         } & \small{34770} & \small{5.83} \\
  \small{J20335842+4019411} & \small{20:33:58.417} & \small{+40:19:41.13} & \small{O9:         } & \small{34770} & \small{5.43} \\
  \small{J20321656+4125357} & \small{20:32:16.563} & \small{+41:25:35.67} & \small{O9V         } & \small{34770} & \small{4.79} \\
  \small{J20311833+4121216} & \small{20:31:18.329} & \small{+41:21:21.65} & \small{O9V         } & \small{34770} & \small{5.06} \\
  \small{J20340486+4105129} & \small{20:34:04.851} & \small{+41:05:11.76} & \small{O9V         } &\small{ 34770} & \small{4.59} \\
  \small{J20311055+4131535} & \small{20:31:10.543} & \small{+41:31:53.53} & \small{O9V         } & \small{34770} & \small{5.14} \\
  \small{J20332101+4117401} & \small{20:33:21.016} & \small{+41:17:40.11} & \small{O9V         } & \small{34770} & \small{4.63} \\
  \small{J20331571+4120172} & \small{20:33:15.685} & \small{+41:20:18.75} & \small{O9V         } & \small{34770} & \small{4.60} \\
  \small{J20301839+4053466} & \small{20:30:18.391} & \small{+40:53:46.56} & \small{O9V         } & \small{34770} & \small{4.99} \\
  \small{J20314965+4128265} & \small{20:31:49.658} & \small{+41:28:26.50} & \small{O9III       } & \small{33120} & \small{4.53} \\
  \small{J20345606+4038179} & \small{20:34:56.057} & \small{+40:38:17.92} & \small{O9.7Iab     } & \small{29922} & \small{5.33} \\
  \small{J20305772+4109575} & \small{20:30:57.727} & \small{+41:09:57.51} & \small{O9.5V       } & \small{33960} & \small{4.85} \\
  \small{J20340601+4108090} & \small{20:34:06.017} & \small{+41:08:09.13} & \small{O9.5V       } & \small{33960} & \small{4.87} \\
  \small{J20335952+4117354} & \small{20:33:59.527} & \small{+41:17:35.46} & \small{O9.5V       } & \small{33960} & \small{4.91} \\
  \small{J20341605+4102196} & \small{20:34:16.046} & \small{+41:02:19.59} & \small{O9.5V       } & \small{33960} & \small{4.73} \\
  \small{J20272428+4115458} & \small{20:27:24.282} & \small{+41:15:45.82} & \small{O9.5V       } & \small{33960} & \small{4.51} \\
  \small{J20293480+4120089} & \small{20:29:34.798} & \small{+41:20:08.93} & \small{O9.5V       } & \small{33960} & \small{4.76} \\
  \small{J20323033+4034332} & \small{20:32:30.310} & \small{+40:34:33.22} & \small{O9.5IV      } & \small{33067} & \small{5.25} \\
  \small{J20333700+4116113} & \small{20:33:36.994} & \small{+41:16:11.31} & \small{O9.5IV      } & \small{33067} & \small{4.88} \\
  \small{J20334610+4133010} & \small{20:33:46.112} & \small{+41:33:01.00} & \small{O9.5Ia      } & \small{30250} & \small{5.71} \\  
\hline
		\end{tabular}		

\end{table*}

 \begin{table*}[ht!]
  \centering  		
	\begin{tabular}{lccccc}
	\hline   
	\hline \\[-1.5ex]
   	\small{Object}& \small{RA (hhmmss)}& \small{Dec ($^{\circ}$ $^\prime$ $^{\prime\prime}$)}& \small{SpT} &\small{$T_{\rm eff}$ (K)} &\small{log ($L/L_{\odot}$)}  \\
   \hline \\[-1.5ex]     
  \small{J20325964+4115146} & \small{20:32:59.633} & \small{+41:15:14.66} & \small{O9.7III   } & \small{31797} & \small{4.80} \\
  \small{J20283039+4105290} & \small{20:28:30.385} & \small{+41:05:29.04} & \small{OC9.7Ia   } & \small{29922} & \small{5.52} \\
  \small{J20302730+4113253} & \small{20:30:27.300} & \small{+41:13:25.13} & \small{Ofpe      } & \small{38612} & \small{5.79} \\
  \small{J20281547+4038196} & \small{20:28:15.471} & \small{+40:38:19.81} & \small{B0V:      } & \small{32816} & \small{4.79} \\
  \small{J20323951+4052475} & \small{20:32:39.507} & \small{+40:52:47.46} & \small{B0:V:     } & \small{32816} & \small{5.19} \\
  \small{J20331050+4122224} & \small{20:33:10.502} & \small{+41:22:22.44} & \small{B0V       } & \small{32816} & \small{4.45} \\
  \small{J20305552+4109575} & \small{20:30:55.516} & \small{+40:54:54.03} & \small{B0V       } & \small{32816} & \small{4.66} \\
  \small{J20295701+4109538} & \small{20:29:57.010} & \small{+41:09:53.84} & \small{B0V       } & \small{32816} & \small{4.64} \\
  \small{J20305111+4120218} & \small{20:30:51.115} & \small{+41:20:21.78} & \small{B0V       } & \small{32816} & \small{4.53} \\
  \small{J20331130+4042337} & \small{20:33:11.300} & \small{+40:42:33.73} & \small{B0:III:   } & \small{30308} & \small{4.72} \\
  \small{J20333821+4041064} & \small{20:33:38.213} & \small{+40:41:06.35} & \small{B0Ia      } & \small{27800} & \small{5.64} \\
  \small{J20344471+4051465} & \small{20:34:44.716} & \small{+40:51:46.73} & \small{B0Ia      } & \small{27800} & \small{5.59} \\
  \small{J20323904+4100078} & \small{20:32:39.057} & \small{+41:00:07.78} & \small{B0Ia      } & \small{27800} & \small{5.79} \\
  \small{J20322774+4128522} & \small{20:32:27.738} & \small{+41:28:52.26} & \small{B0Ib      } & \small{27800} & \small{4.24} \\
  \small{J20345878+4136174} & \small{20:34:58.781} & \small{+41:36:17.35} & \small{B0Ib(n)sb } & \small{27800} & \small{5.31} \\
  \small{J20333822+4053412} & \small{20:33:38.218} & \small{+40:53:41.19} & \small{B0Ib      } & \small{27800} & \small{4.94} \\
  \small{J20333910+4119258} & \small{20:33:39.102} & \small{+41:19:25.98} & \small{B0Iab     } & \small{27800} & \small{5.23} \\
  \small{J20323498+4052390} & \small{20:32:34.848} & \small{+40:52:39.46} & \small{B0.2V     } & \small{31906} & \small{4.58} \\
  \small{J20292449+4052599} & \small{20:29:24.485} & \small{+40:52:59.85} & \small{B0.2IV    } & \small{30588} & \small{4.71} \\
  \small{J20321568+4046170} & \small{20:32:15.679} & \small{+40:46:17.00} & \small{B0.2IV    } & \small{30588} & \small{4.39} \\
  \small{J20294666+4105083} & \small{20:29:46.672} & \small{+41:05:08.32} & \small{B0.5V(n)sb} & \small{30288} & \small{4.50} \\
  \small{J20282772+4104018} & \small{20:28:27.723} & \small{+41:04:01.80} & \small{B0.5V     } & \small{30288} & \small{4.47} \\
  \small{J20314605+4043246} & \small{20:31:46.053} & \small{+40:43:24.61} & \small{B0.5IV    } & \small{28969} & \small{4.55} \\
  \small{J20331870+4059379} & \small{20:33:18.696} & \small{+40:59:37.92} & \small{B0.5IIIe  } & \small{27651} & \small{4.64} \\
  \small{J20303970+4108489} & \small{20:30:39.701} & \small{+41:08:48.80} & \small{B0.7Ib    } & \small{24014} & \small{5.30} \\
  \small{J20294060+4109585} & \small{20:29:40.601} & \small{+41:09:58.54} & \small{B1[e]     } & \small{27173} & \small{4.27} \\
  \small{J20313338+4122490} & \small{20:31:33.378} & \small{+41:22:49.02} & \small{B1V       } & \small{27173} & \small{4.25} \\
  \small{J20340435+4108078} & \small{20:34:04.349} & \small{+41:08:07.91} & \small{B1V       } & \small{27173} & \small{4.27} \\
  \small{J20303833+4010538} & \small{20:30:38.329} & \small{+40:10:53.84} & \small{B1V       } & \small{27173} & \small{4.34} \\
  \small{J20293473+4020381} & \small{20:29:34.728} & \small{+40:20:38.09} & \small{B1V       } & \small{27173} & \small{4.38} \\
  \small{J20273982+4040384} & \small{20:27:39.821} & \small{+40:40:38.35} & \small{B1V       } & \small{27173} & \small{4.28} \\
  \small{J20303297+4044024} & \small{20:30:32.965} & \small{+40:44:02.41} & \small{B1V       } & \small{27173} & \small{4.31} \\
  \small{J20310464+4030568} & \small{20:31:04.659} & \small{+40:30:56.93} & \small{B1III:e   } & \small{24903} & \small{5.95} \\
  \small{J20334783+4120415} & \small{20:33:47.831} & \small{+41:20:41.37} & \small{B1III     } & \small{24903} & \small{4.80} \\
  \small{J20281539+4044046} & \small{20:28:15.392} & \small{+40:44:04.57} & \small{B1III     } & \small{24903} & \small{4.58} \\
  \small{J20310700+4035537} & \small{20:31:07.003} & \small{+40:35:53.73} & \small{B1III     } & \small{24903} & \small{4.08} \\
  \small{J20314885+4038001} & \small{20:31:48.848} & \small{+40:38:00.05} & \small{B1II      } & \small{23767} & \small{4.04} \\
  \small{J20333078+4115226} & \small{20:33:30.791} & \small{+41:15:22.70} & \small{B1I       } & \small{22632} & \small{5.41} \\
  \small{J20312203+4131284} & \small{20:31:22.026} & \small{+41:31:28.40} & \small{B1Ib:     } & \small{22632} & \small{4.45} \\
  \small{J20322734+4055184} & \small{20:32:27.339} & \small{+40:55:18.25} & \small{B2V       } & \small{21092} & \small{4.39} \\
  \small{J20312210+4112029} & \small{20:31:22.101} & \small{+41:12:02.87} & \small{B2V       } & \small{21092} & \small{4.10} \\
  \small{J20354703+4053012} & \small{20:35:47.026} & \small{+40:53:01.17} & \small{B2V       } & \small{21092} & \small{3.91} \\
  \small{J20284657+4107069} & \small{20:28:46.566} & \small{+41:07:06.86} & \small{B2II      } & \small{19482} & \small{3.83} \\
  \small{J20320689+4117570} & \small{20:32:06.877} & \small{+41:17:56.97} & \small{B3V       } & \small{18546} & \small{3.91} \\
    \cline{1-1}  \\[-1.8ex]  
\small{\textit{    Cygnus OB9  }} & \small{} & \small{}  & \small{} & \small{}& \small{}   \\  
    \cline{1-1}   \\[-1.5ex]  
  \small{HD229196}          & \small{20:23:10.784} & \small{+40:52:29.85} & \small{O5     } & \small{41250} & \small{5.89} \\
  \small{J20223777+4140292} & \small{20:22:37.766} & \small{+41:40:29.23} & \small{O5If   } & \small{37630} & \small{5.48} \\
  \small{BD+394177}         & \small{20:25:22.122} & \small{+40:13:01.09} & \small{O6.5   } & \small{38820} & \small{5.55} \\
  \small{HD229250}          & \small{20:24:11.733} & \small{+39:40:41.54} & \small{O7     } & \small{38010} & \small{5.44} \\
  \small{BD+394168}         & \small{20:24:21.475} & \small{+39:46:03.90} & \small{O7     } & \small{38010} & \small{5.48} \\
  \small{HD229202}          & \small{20:23:22.840} & \small{+40:09:22.53} & \small{O8V:   } & \small{36390} & \small{5.14} \\
  \small{BD+404159}         & \small{20:25:06.521} & \small{+40:35:49.78} & \small{O9V    } & \small{34770} & \small{4.61} \\
  \small{BD+404148}         & \small{20:23:14.549} & \small{+40:45:19.07} & \small{O9.5:V } & \small{33960} & \small{4.88} \\
  \small{J20194916+4052090} & \small{20:19:49.156} & \small{+40:52:08.99} & \small{O9.5V  } & \small{33960} & \small{4.86} \\
  \small{J20190610+4037004} & \small{20:19:06.102} & \small{+40:37:00.39} & \small{O9.7Iab} & \small{29922} & \small{4.74} \\
  \small{HD193945}          & \small{20:21:25.823} & \small{+41:11:39.56} & \small{B0Vnn  } & \small{32816} & \small{4.92} \\
  \small{BD+384058}         & \small{20:23:28.531} & \small{+39:20:59.05} & \small{B0V    } & \small{32816} & \small{4.83} \\
  \small{LSII+4032}         & \small{20:25:28.893} & \small{+40:12:54.13} & \small{B0III  } & \small{30308} & \small{4.36} \\
  \small{J20243872+3930301} & \small{20:24:38.720} & \small{+39:30:30.10} & \small{B0I:   } & \small{27800} & \small{4.58} \\
  \small{J20183413+4025045} & \small{20:18:34.130} & \small{+40:25:04.47} & \small{B0.2IV } & \small{30588} & \small{4.86} \\
  \small{NGC6910-14}        & \small{20:23:07.575} & \small{+40:46:08.87} & \small{B0.5V  } & \small{30288} & \small{4.56} \\  
\hline
		\end{tabular}		
\end{table*}

 \begin{table*}[ht!]
   	\centering	
	\begin{tabular}{lccccc}
	\hline   
	\hline \\[-1.5ex]
   	\small{Object}& \small{RA (hhmmss)}& \small{Dec ($^{\circ}$ $^\prime$ $^{\prime\prime}$)}& \small{SpT} &\small{$T_{\rm eff}$ (K)} & \small{log ($L/L_{\odot}$)}  \\
   \hline  \\[-1.5ex]      
  \small{J20240515+4046035} & \small{20:24:05.154} & \small{+40:46:03.51} & \small{B0.5V     } & \small{30288} & \small{4.60}\\
  \small{HD194092}          & \small{20:22:05.443} & \small{+40:59:08.17} & \small{B0.5III   } & \small{27651} & \small{4.39}\\
  \small{HD228882}          & \small{20:18:57.784} & \small{+40:42:18.52} & \small{B0.5Ia    } & \small{25014} & \small{5.25}\\
  \small{HD228929}          & \small{20:19:36.542} & \small{+39:54:41.80} & \small{B0.5Ib    } & \small{25014} & \small{5.13}\\
  \small{J20234624+3937078} & \small{20:23:46.238} & \small{+39:37:07.83} & \small{B0.7IV    } & \small{27818} & \small{4.54}\\
  \small{J20241767+3920326} & \small{20:24:17.666} & \small{+39:20:32.56} & \small{B1V       } & \small{27173} & \small{4.25}\\
  \small{J20214868+4043005} & \small{20:21:48.682} & \small{+40:43:00.45} & \small{B1V       } & \small{27173} & \small{4.32}\\
  \small{HD228919}          & \small{20:19:27.908} & \small{+40:27:42.09} & \small{B1IV      } & \small{26038} & \small{4.34}\\
  \small{J20233375+4045199} & \small{20:23:33.752} & \small{+40:45:19.93} & \small{B1III     } & \small{24903} & \small{4.03}\\
  \small{J20223944+3935420} & \small{20:22:39.442} & \small{+39:35:42.02} & \small{B1III     } & \small{24903} & \small{4.42}\\
  \small{J20220454+4042487} & \small{20:22:04.541} & \small{+40:42:48.73} & \small{B1III     } & \small{24903} & \small{4.28}\\
  \small{J20215593+4110129} & \small{20:21:55.930} & \small{+41:10:12.92} & \small{B1III     } & \small{24903} & \small{4.03}\\
  \small{J20220879+3958161} & \small{20:22:08.793} & \small{+39:58:16.07} & \small{B1II      } & \small{23767} & \small{4.38}\\
  \small{J20214410+4012529} & \small{20:21:44.103} & \small{+40:12:52.91} & \small{B1Ia      } & \small{22632} & \small{5.18}\\
  \small{J20215160+3959496} & \small{20:21:51.600} & \small{+39:59:49.61} & \small{B1Ib      } & \small{22632} & \small{3.94}\\
  \small{J20203933+4031176} & \small{20:20:39.334} & \small{+40:31:17.64} & \small{B1.5V     } & \small{24132} & \small{4.08}\\
  \small{J20204933+4033027} & \small{20:20:49.333} & \small{+40:33:02.73} & \small{B1.5V     } & \small{24132} & \small{4.15}\\
  \small{HD228911}          & \small{20:19:21.712} & \small{+40:53:16.46} & \small{B2        } & \small{21092} & \small{4.25}\\
  \small{HD194194}          & \small{20:22:44.760} & \small{+40:42:52.63} & \small{B2III     } & \small{20019} & \small{3.91}\\
  \small{J20211677+4023162} & \small{20:21:16.773} & \small{+40:23:16.19} & \small{B2III     } & \small{20019} & \small{3.94}\\
  \small{J20250591+4020124} & \small{20:25:05.912} & \small{+40:20:12.44} & \small{B2III     } & \small{20019} & \small{3.75}\\
  \small{HD228928}          & \small{20:19:32.709} & \small{+40:39:13.75} & \small{B2Ib:nn   } & \small{18945} & \small{4.38}\\
  \small{HD228941}          & \small{20:19:40.169} & \small{+40:53:19.19} & \small{B3        } & \small{18546} & \small{3.98}\\
  \small{BD+404146}         & \small{20:23:10.464} & \small{+40:45:52.34} & \small{B3        } & \small{18546} & \small{4.35}\\
  \small{NGC6910-16}        & \small{20:23:07.301} & \small{+40:46:55.25} & \small{B3        } & \small{18546} & \small{4.05}\\
  \small{J20215115+3934215} & \small{20:21:51.149} & \small{+39:37:51.47} & \small{B3V       } & \small{18546} & \small{4.18}\\
  \small{J20221729+3946035} & \small{20:22:17.286} & \small{+39:34:21.50} & \small{B5Ia      } & \small{14012} & \small{4.01}\\
  \small{HD228821}          & \small{20:18:04.930} & \small{+40:06:06.80} & \small{B8        } & \small{12449} & \small{3.03}\\
  \small{HD193426}          & \small{20:18:39.749} & \small{+40:13:36.89} & \small{B9Ia      } & \small{11457} & \small{4.85}\\
    \cline{1-1}   \\[-1.8ex]      
\small{\textit{  Boundaries}} & \small{} & \small{}  & \small{} & \small{}& \small{}  \\  
     \cline{1-1}   \\[-1.5ex]        
  \small{BD+433654}         & \small{20:33:36.079} & \small{+43:59:07.38} & \small{O4If     } & \small{39270} & \small{5.87}\\
  \small{LSII+3953}         & \small{20:27:17.572} & \small{+39:44:32.60} & \small{O7V:     } & \small{38010} & \small{5.07}\\
  \small{BD+42376}0         & \small{20:28:40.812} & \small{+43:08:58.46} & \small{O8.5V    } & \small{35580} & \small{4.96}\\
  \small{BD+423835}         & \small{20:42:06.863} & \small{+43:11:03.72} & \small{O9p...   } & \small{34770} & \small{5.25}\\
  \small{J20342894+4156171} & \small{20:34:28.941} & \small{+41:56:17.09} & \small{O9V      } & \small{34770} & \small{4.86}\\
  \small{J20462826+4223417} & \small{20:46:28.255} & \small{+42:23:41.74} & \small{O9V      } & \small{34770} & \small{4.88}\\
  \small{J20272553+3929246} & \small{20:27:25.529} & \small{+39:29:24.58} & \small{O9.5V    } & \small{33960} & \small{4.88}\\
  \small{J20325571+4307583} & \small{20:32:55.713} & \small{+43:07:58.26} & \small{O9.5V    } & \small{33960} & \small{4.97}\\
  \small{J20310838+4202422} & \small{20:31:08.376} & \small{+42:02:42.25} & \small{O9.7II   } & \small{30859} & \small{5.15}\\
  \small{HD199021}          & \small{20:52:53.207} & \small{+42:36:27.87} & \small{B0V      } & \small{32816} & \small{5.05}\\
  \small{J20382040+4156563} & \small{20:38:20.413} & \small{+41:56:56.51} & \small{B0II     } & \small{29054} & \small{5.29}\\
  \small{J20385918+4202395} & \small{20:38:59.181} & \small{+42:02:39.45} & \small{B0Ib     } & \small{27800} & \small{4.67}\\
  \small{J20294195+3859342} & \small{20:29:41.952} & \small{+38:59:34.16} & \small{B0.2V    } & \small{31906} & \small{4.61}\\
  \small{J20352227+4355305} & \small{20:35:22.266} & \small{+43:55:30.46} & \small{B0.2IV   } & \small{30588} & \small{4.90}\\
  \small{BD+413794}         & \small{20:32:02.204} & \small{+42:12:26.15} & \small{B0.2III  } & \small{29270} & \small{4.97}\\
  \small{HD194839}          & \small{20:26:21.545} & \small{+41:22:45.65} & \small{B0.5Iae  } & \small{25014} & \small{5.73}\\
  \small{J20313693+4201218} & \small{20:31:36.921} & \small{+42:01:21.79} & \small{B0.7Ib   } & \small{24014} & \small{4.96}\\
  \small{J20301273+3904216} & \small{20:30:12.732} & \small{+39:04:21.59} & \small{B1V      } & \small{27173} & \small{4.24}\\
  \small{LSIII+4217}        & \small{20:35:10.623} & \small{+42:20:22.83} & \small{B1III    } & \small{24903} & \small{4.69}\\
  \small{J20374323+4232334} & \small{20:37:43.232} & \small{+42:32:33.40} & \small{B1III    } & \small{24903} & \small{4.13}\\
  \small{J20314215+4225532} & \small{20:31:42.151} & \small{+42:25:53.26} & \small{B1Ib     } & \small{22632} & \small{5.67}\\
  \small{BD+373976}         & \small{20:33:49.752} & \small{+38:17:00.06} & \small{B1.5Vn   } & \small{24132} & \small{4.17}\\
  \small{J20313853+4152585} & \small{20:31:38.532} & \small{+41:52:58.46} & \small{B1.5V    } & \small{24132} & \small{4.07}\\
  \small{J20312725+4304227} & \small{20:31:27.253} & \small{+43:04:22.67} & \small{B1.5V    } & \small{24132} & \small{4.33}\\
  \small{BD+394189}         & \small{20:26:20.922} & \small{+39:40:10.06} & \small{B2p?e?   } & \small{21092} & \small{4.85}\\
  \small{BD+413762}         & \small{20:28:15.212} & \small{+42:25:39.14} & \small{B2V      } & \small{21092} & \small{4.32}\\
  \small{J20452110+4223514} & \small{20:45:21.103} & \small{+42:23:51.37} & \small{B2V      } & \small{21092} & \small{3.99}\\
  \small{J20264025+4233221} & \small{20:26:40.251} & \small{+42:33:22.09} & \small{B2II     } & \small{19482} & \small{3.67}\\
  \small{HD196489}          & \small{20:36:24.259} & \small{+39:11:40.70} & \small{B3V      } & \small{18546} & \small{3.92}\\
  \small{J20462289+4212311} & \small{20:46:22.892} & \small{+42:12:31.07} & \small{B3V      } & \small{18546} & \small{4.06}\\
  \small{HD194779}          & \small{20:25:55.077} & \small{+41:20:11.73} & \small{B3II     } & \small{16983} & \small{4.12}\\
  \small{BD+384098}         & \small{20:27:33.010} & \small{+38:46:19.62} & \small{B9Ib     } & \small{11457} & \small{3.92}\\
\hline
		\end{tabular}		
\end{table*}

\section{Candidate spectra}\label{appb1}
\begin{figure*}[ht!]
  \centering
\caption{ Spectra of the 61 OB candidate stars where dotted vertical lines indicate H, He, Si and Mg lines in the wavelength range.}  
\label{figb1}
\includegraphics[width=18cm ]{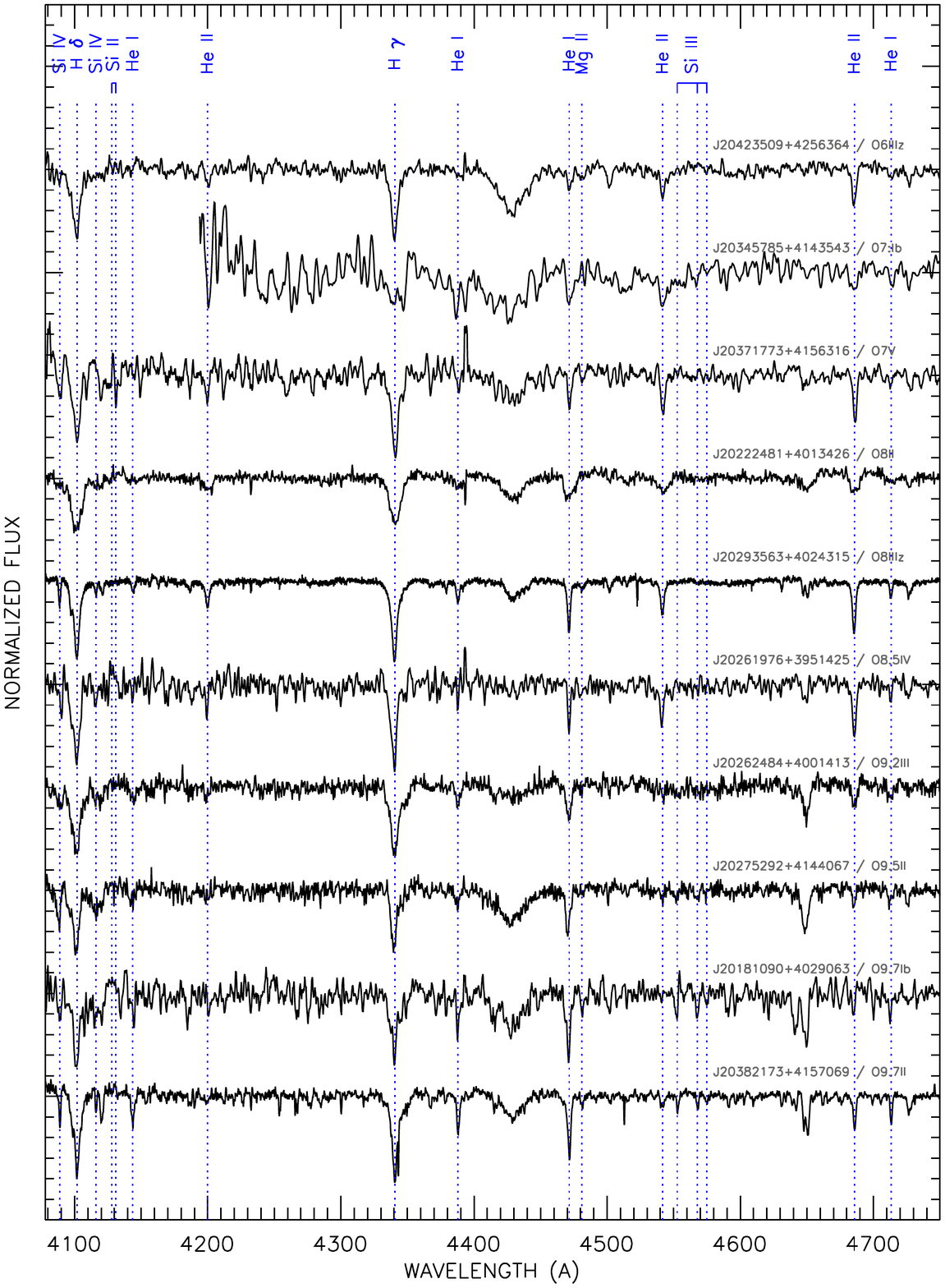} 
\end{figure*}

In Fig.~\ref{figb1} are plotted the normalized spectra of the 61 OB candidates from the list of \cite{com12}. The spectra have been corrected for the stellar radial velocities and diagnostic lines are also indicated.

  \begin{figure*}[ht!]
\centering
\includegraphics[width=18cm ]{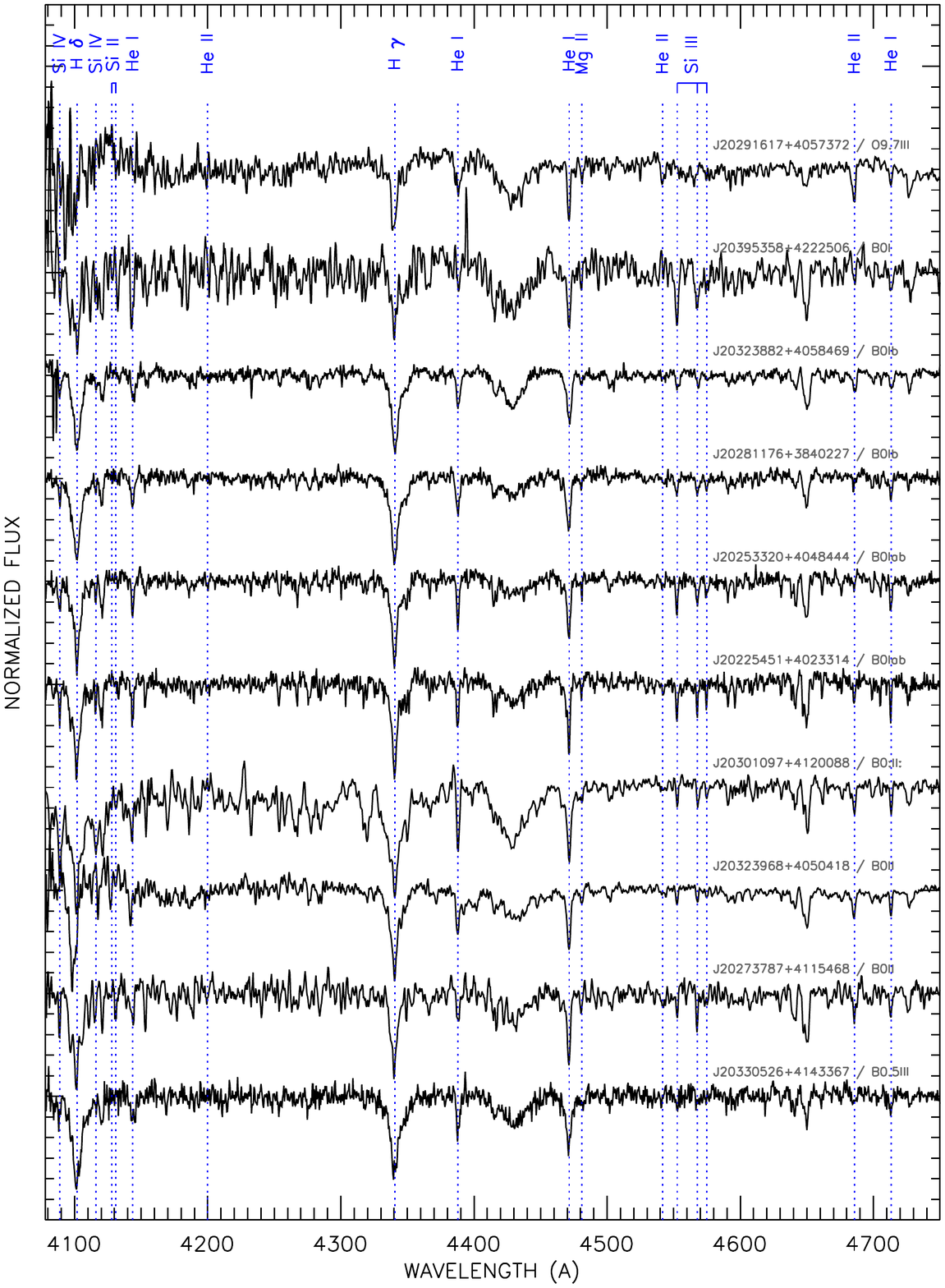} 
\end{figure*}

  \begin{figure*}[ht!]
\centering
\includegraphics[width=18cm ]{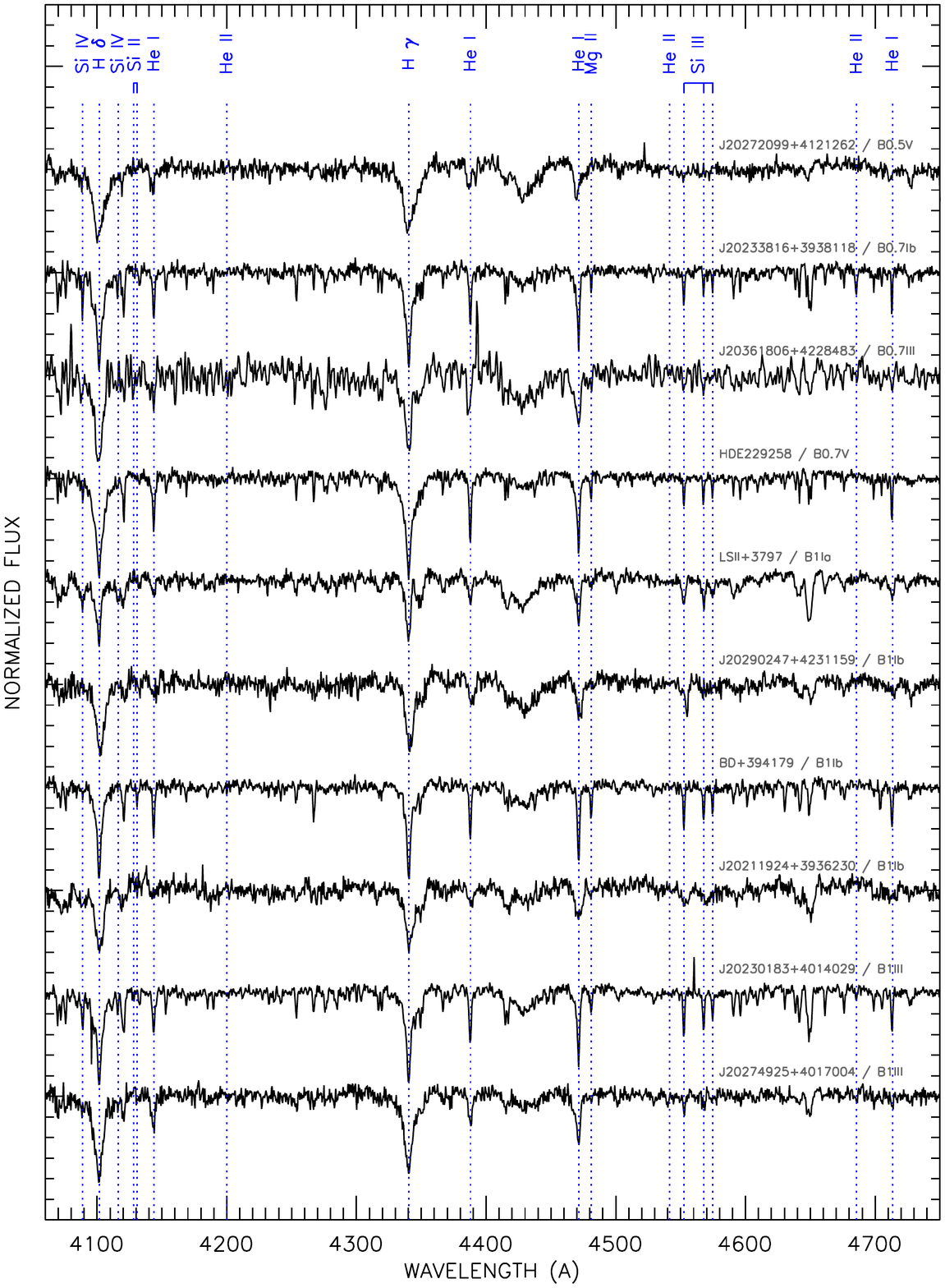} 
\end{figure*}

  \begin{figure*}[ht!]
\centering
\includegraphics[width=18cm ]{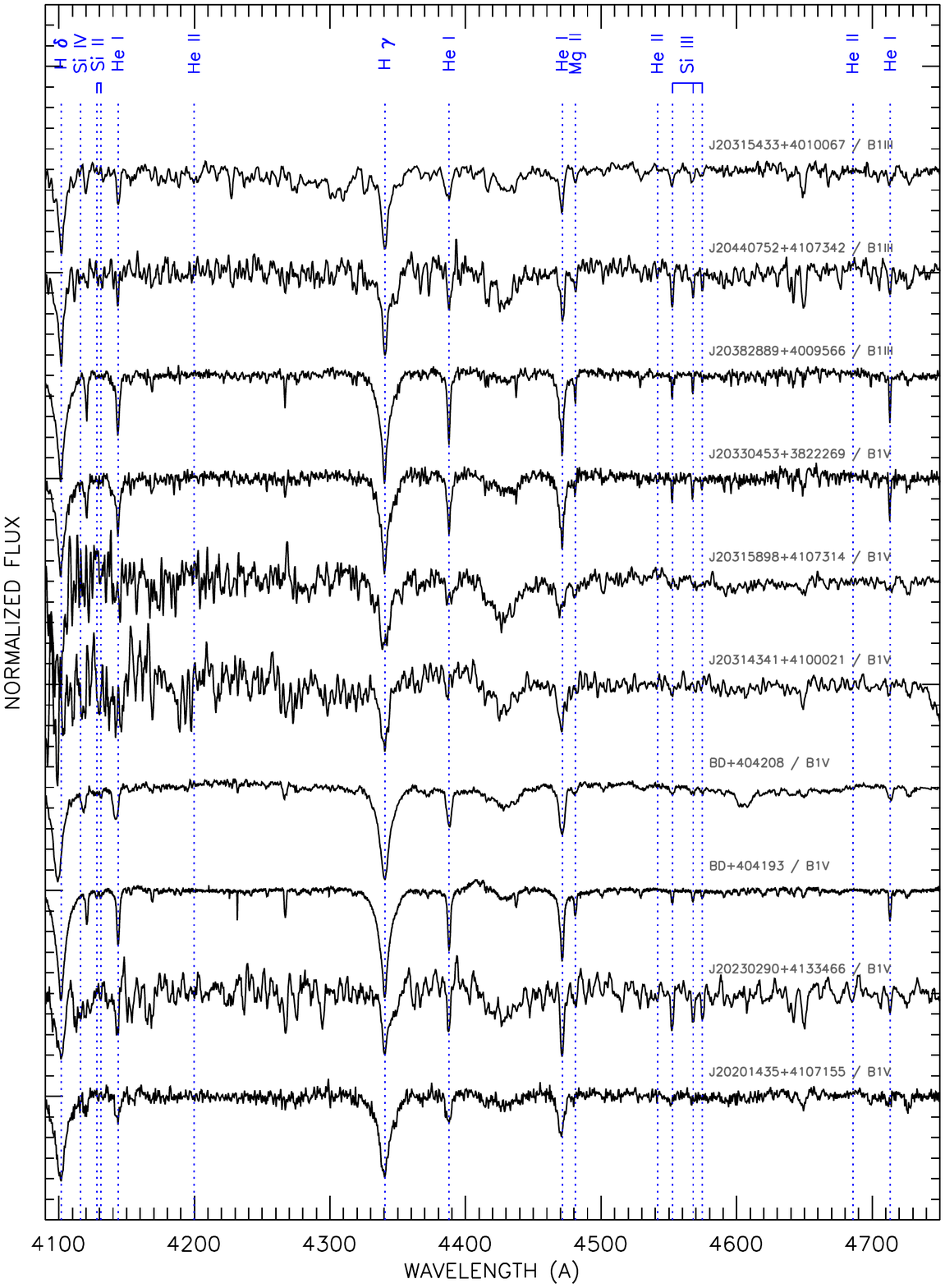} 
\end{figure*}

  \begin{figure*}[ht!]
\centering
\includegraphics[width=18cm ]{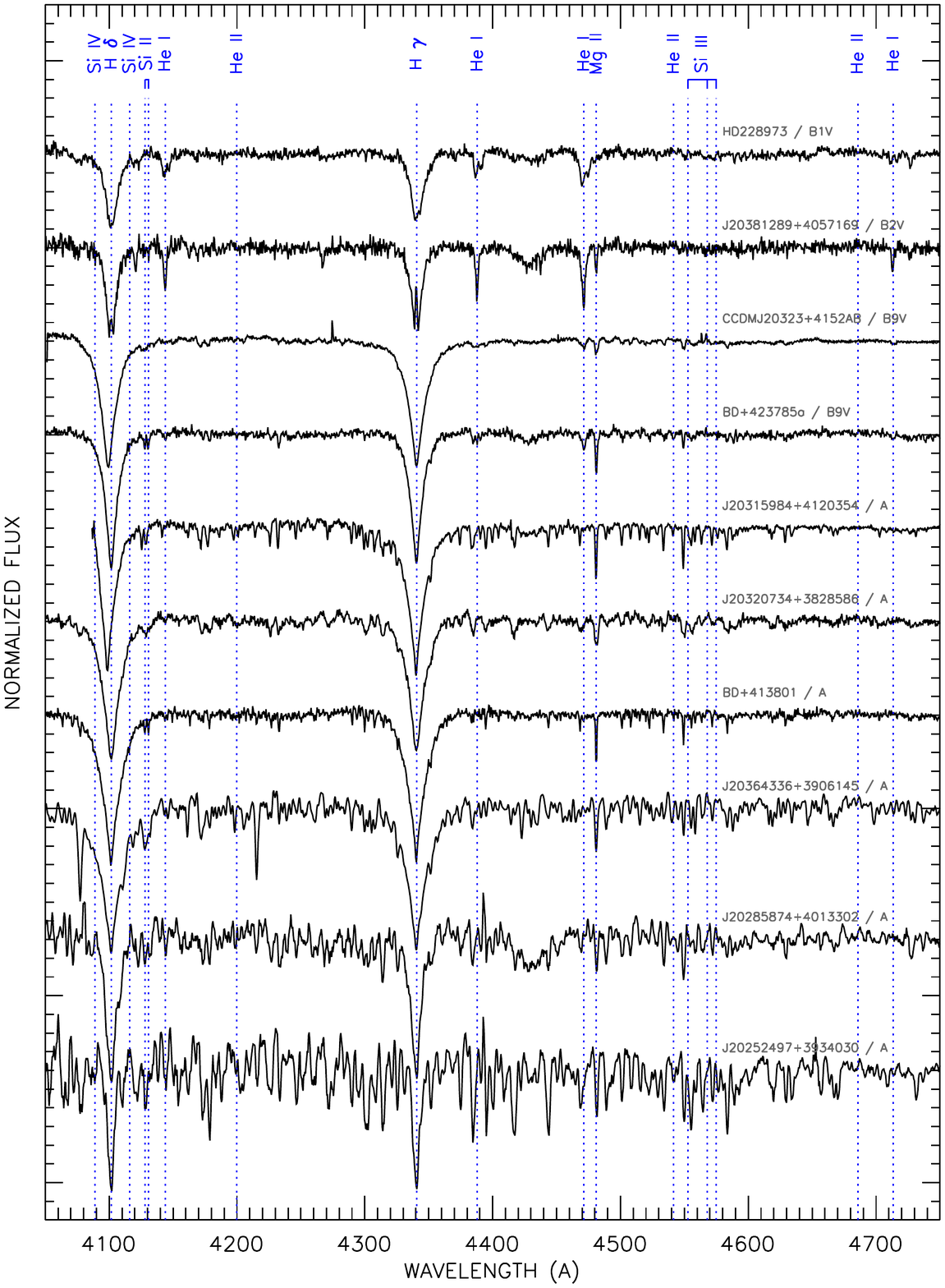} 
\end{figure*}

  \begin{figure*}[ht!]
\centering
\includegraphics[width=18cm ]{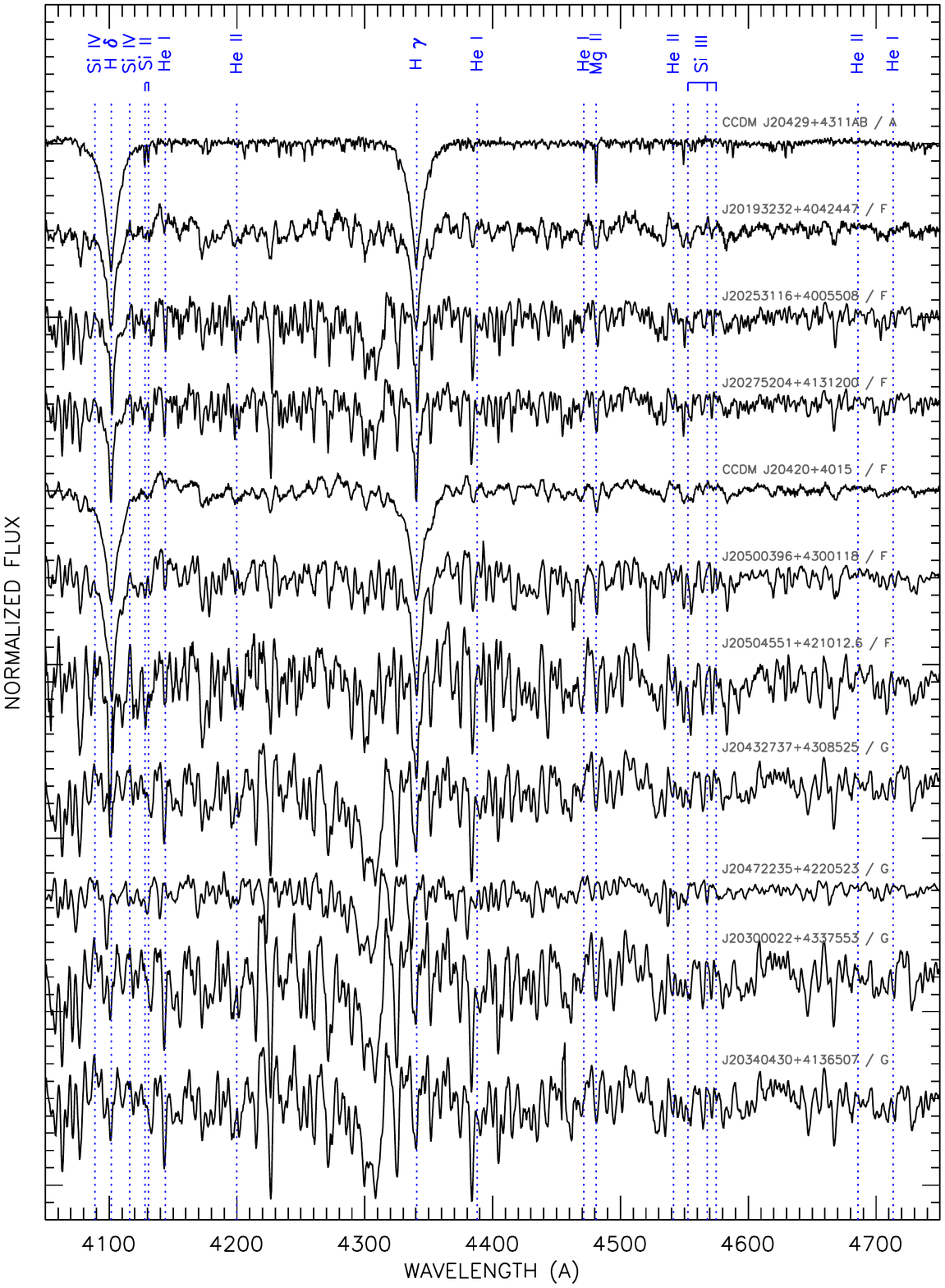} 
\end{figure*}

\end{appendix}

\end{document}